\documentclass[PAPER, cernpreprint=true, atlasdraft=false, USenglish, orcidlogo, texmf, orcidlogo]{atlasdoc}
\usepackage[subfigure,rotating]{atlaspackage}
\usepackage{atlasbiblatex}
\usepackage{makecell}

\usepackage{atlasphysics}
\usepackage{comment}
\ifthenelse{\boolean{AtlasDraft}}{\usepackage[output=true, shift=true]{atlasdoc}}{}

\graphicspath{{logos/}{figures/}}

\usepackage{ANA-STDM-2019-29-PAPER-defs}

\AtlasTitle{Evidence for longitudinally polarized $W$ bosons in the electroweak production of same-sign $W$ boson pairs in association with two jets in pp collisions at $\sqrt{s}=13$ TeV  with the ATLAS detector}

\AtlasAbstract{%
This Letter reports the first evidence of electroweak production of same-sign $W$ boson pairs where at least one of the $W$ bosons is longitudinally polarized and the most stringent constraint to date for the production of two longitudinally polarized same-sign $W$ bosons. The data set used corresponds to an integrated luminosity of 140~fb$^{-1}$ of proton--proton collisions at a center-of-mass energy of 13 TeV, collected with the ATLAS detector during Run 2 of the Large Hadron Collider. The study is performed in final states including two same-sign leptons (electrons or muons), missing transverse momentum, and at least two jets with a large invariant mass and a large rapidity difference. Two independent fits are performed targeting the production of same-sign $W$ bosons with at least one, or two longitudinally polarized $W$ bosons. The observed (expected) significance of the production with at least one longitudinally polarized $W$ boson is 3.3 (4.0) standard deviations. An observed (expected) 95\% confidence level upper limit of 0.45 (0.70) fb is reported on the fiducial production cross section of two longitudinally polarized same-sign $W$ bosons.
}

\AtlasRefCode{STDM-2019-29}

\PreprintIdNumber{CERN-EP-2025-048}

\AtlasJournal{\PRL}
\AtlasJournalRef{\PRL 135 (2025) 111802}
\AtlasDOI{10.1103/bpln-ccql}

\hypersetup{pdftitle={ATLAS document},pdfauthor={The ATLAS Collaboration}}

\begin{document}

\maketitle

The longitudinal polarization states of the $W$ and $Z$ bosons are generated by the Goldstone bosons from spontaneous electroweak (EW) symmetry breaking via the Higgs mechanism~\cite{Higgs:1964ia, Higgs:1964pj, Englert:1964et, Guralnik:1964eu, Kibble:1967sv} within the Standard Model (SM) of particle physics. The characterization of the Higgs boson discovered in 2012~\cite{HIGG-2012-27, CMS-HIG-12-028, HIGG-2021-23, CMS-HIG-22-001} at the Large Hadron Collider~\cite{Evans:2008zzb} (LHC) is an important avenue of research for physics beyond the SM. Polarized production of $W$ and $Z$ boson pairs via vector-boson-scattering (VBS) processes provides a sensitive test of the EW symmetry breaking mechanism. In particular, couplings of longitudinally polarized $W$ and $Z$ bosons to the Higgs boson prevent the divergence of tree-level VBS amplitudes at high energies when the two outgoing bosons are longitudinally polarized, restoring the unitarity at the~\TeV scale~\cite{Lee:1977yc,Lee:1977eg,Alboteanu:2008my}. This Letter reports studies of the EW production of polarized same-sign $W$ boson pairs (\ssww). The first evidence of production with at least one of the $W$ bosons longitudinally polarized is reported along with the most stringent constraint to date on the production of two longitudinally polarized same-sign $W$ bosons. The \ssww candidates are selected using the leptonic decay modes of the $W$ bosons into electrons or muons, $\ssww \rightarrow \ell^\pm\nu \ell'^\pm\nu$, where the combination of $\ell$ and $\ell^{\prime}$ can be $e^\pm e^\pm$, $e^\pm \mu^\pm$ or $\mu^\pm \mu^\pm$. The data set corresponds to an integrated luminosity of $140.1 \pm 1.2$~fb$^{-1}$~\cite{DAPR-2021-01,LUCID2} of 13~\TeV proton--proton ($\pp$) collisions, collected with the ATLAS detector~\cite{PERF-2007-01} during Run 2 of the LHC (2015--2018).

\begin{figure}[!bp]
\centering
\subfigure[]{\includegraphics[width=.24\textwidth]{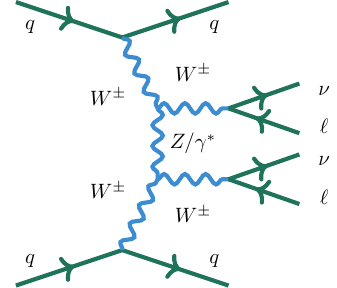}}
\subfigure[]{\includegraphics[width=.24\textwidth]{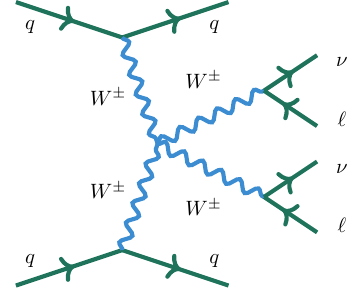}}
\subfigure[]{\includegraphics[width=.24\textwidth]{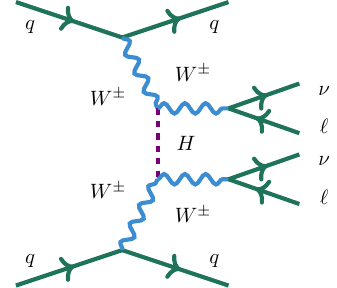}}
\subfigure[]{\includegraphics[width=.24\textwidth]{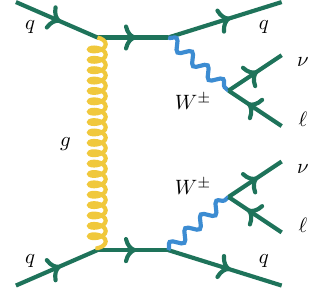}}
\caption{Representative Feynman diagrams for a VBS EW \sswwjj production that either include (a) triple-gauge-boson vertices, (b) a quartic gauge boson vertex, or (c) the exchange of a Higgs boson, and (d) for a QCD \sswwjj production.}
\label{fig:diagrams_ewk_vbs}
\end{figure}

In $\pp$ collisions, the VBS processes at leading-order (LO) involve two initial quarks each of which radiates a gauge boson. The process is characterized by the subsequent interaction and decay of the two gauge bosons and by the presence of two jets ($j$), typically close to the beam direction. In the VBS production of same sign $W$ boson pairs, the final state consists of two leptons with the same electric charge, two neutrinos, and two jets with a large rapidity separation. The EW production of \ssww (EW \sswwjj) involves only EW interaction vertices, of which the VBS processes constitute an important component. The production of \sswwjj involving strong interaction vertices at LO (QCD \sswwjj) is relatively small as quark--gluon and gluon--gluon initiated diagrams are not present at LO. Representative Feynman diagrams for EW \sswwjj and QCD \sswwjj processes are shown in Figure~\ref{fig:diagrams_ewk_vbs}. The ratio of the EW \sswwjj to the QCD \sswwjj production cross sections is of the order of five in the fiducial region of this analysis.

The first experimental observation of the unpolarized EW \sswwjj production were reported by the CMS and ATLAS Collaborations at $\sqrt{s}=13$~\TeV using a partial LHC Run~2 data corresponding to an integrated luminosity of approximately 36~fb$^{-1}$~\cite{CMS-SMP-17-004,STDM-2017-06}. Measurements of cross sections for the production of unpolarized EW \sswwjj process were recently reported by the ATLAS and CMS Collaborations using the full Run~2 data set, corresponding to an integrated luminosity of approximately $140$~fb$^{-1}$~\cite{CMS-SMP-19-012,STDM-2018-32}. The CMS Collaboration published first studies of the polarized EW \sswwjj production, reporting an observed (expected) significance of 2.3 (3.1) standard deviations for the production with at least one longitudinally polarized $W$ boson~\cite{CMS-SMP-20-006}.

The polarized EW \sswwjj production is classified into three contributions, \sswwll, \sswwlt, and \sswwtt, where $W_{\mathrm{L}}^{\pm}$ and $W_{\mathrm{T}}^{\pm}$ denote longitudinally and transversely polarized $W$ bosons, respectively.
The differences in the emission process from the initial-state quarks, scattering, and decay of the $W_{\mathrm{L}}^{\pm}$ and $W_{\mathrm{T}}^{\pm}$ bosons result in kinematic differences for the final state leptons and jets~\cite{Doroba:2012pd}.  The limited data sample size does not allow for a precise simultaneous measurement of the three contributions. In addition, there are non-trivial correlations between event kinematic variables, and a single kinematic distribution is not sufficient to optimally separate the different polarization contributions. Consequently, dedicated deep neural networks (DNN) are trained independently using kinematic variables to separate the \sswwll contribution from the \sswwlt and \sswwtt contributions (\sswwtx) and to separate the \sswwll and \sswwlt contributions (\sswwlx) from the \sswwtt contribution.

The polarization vectors are defined in the $\ssww$ center-of-mass reference frame. The studies are also performed using the center-of-mass reference frame of the initial-state partons to define the polarization vectors. The choice of the $\ssww$ center-of-mass reference frame leads to a larger contribution of the \sswwll process compared to the center-of-mass reference frame of the initial-state partons or protons~\cite{Ballestrero:2020qgv, Denner:2024tlu}. The expected contributions of the $\sswwll$ and $\sswwlt$ processes, defined using the $\ssww$ center-of-mass reference frame, are about 10\% and 30\% of the overall expected EW $\sswwjj$ production cross section, respectively.  The fractions of the $\sswwll$ ($\sswwlt$) and $\sswwlt$ ($\sswwtt$) processes in the $\sswwlx$ ($\sswwtx$) signal sample are fixed to the SM prediction.

The ATLAS detector~\cite{PERF-2007-01} is a multipurpose particle physics detector with cylindrical geometry\footnote{ATLAS uses a right-handed coordinate system with its origin at the nominal interaction point (IP) in the centre of the detector and the \(z\)-axis along the beam pipe. The \(x\)-axis points from the IP to the centre of the LHC ring, and the \(y\)-axis points upwards. Polar coordinates \((r,\phi)\) are used in the transverse plane, \(\phi\) being the azimuthal angle around the \(z\)-axis.
The pseudorapidity is defined in terms of the polar angle \(\theta\) as \(\eta = -\ln \tan(\theta/2)\) and is equal to the rapidity $ y = \frac{1}{2} \ln \left( \frac{E + p_z}{E - p_z} \right) $ in the relativistic limit. Transverse momentum ($\pt$) is defined relative to the beam axis and is calculated as $p_T = p \sin \theta$ where $p$ is the momentum. Angular distance is measured in units of \(\Delta R \equiv \sqrt{(\Delta y)^{2} + (\Delta\phi)^{2}}\).}. It consists of an inner tracking detector (ID) surrounded by a superconducting solenoid providing a 2 T axial magnetic field, sampling electromagnetic (ECAL) and hadronic (HCAL) calorimeters, and a muon spectrometer (MS) based on three air-core toroidal superconducting magnets with eight coils each. The ATLAS trigger system consists of a hardware-based level-1 trigger followed by a software-based high-level trigger~\cite{TRIG-2016-01}.  An extensive software suite~\cite{SOFT-2022-02} is used in the reconstruction and analysis of real and simulated data, in detector operations, and in the trigger and data acquisition systems of the experiment.

Detailed information about the reconstruction, identification, and calibration of physics objects, as well as the simulation, triggers, and event selection used in this analysis, can be found in Ref.~\cite{STDM-2018-32} which reports the measurements of fiducial and differential cross sections of the unpolarized \sswwjj production. A brief description is given here. The final state probed by this analysis consists in the leptonic decays (electrons or muons) of two $W$ bosons produced in association with two jets. Candidate events used for this analysis are selected with single-lepton ($e$ or $\mu$) triggers~\cite{TRIG-2016-01, TRIG-2018-05, ATL-DAQ-PUB-2019-001}.

Electron candidates are reconstructed from ECAL energy clusters and matched to a track reconstructed in the ID~\cite{EGAM-2018-01}. Electrons are identified using a likelihood function based on shower shape variables in the ECAL, ID track variables, and the quality of the track-cluster matching. Electrons are required to satisfy the ``Tight" likelihood-based identification criteria~\cite{PERF-2017-01} and have to have $p_T>27$~\GeV and $|\eta|<2.47$ excluding the calorimeter transition region $1.37<|\eta|<1.52$. A charge selector tool based on boosted decision trees uses shower shape and track-to-cluster matching variables~\cite{EGAM-2018-01} to reject electron candidates where the charge is likely misidentified. Muons are reconstructed from tracks in the MS, matched to a corresponding track in the ID~\cite{MUON-2018-03}. Muons are identified by a set of
requirements on the number of hits in the diﬀerent ID subdetectors and diﬀerent MS stations, on the track fit properties, and on variables that test the compatibility of the individual measurements in the two detector systems. Muons are required to satisfy the ``medium'' identification selection~\cite{MUON-2018-03} and have $p_T > 27$~\GeV and $|\eta|<2.5$. Electrons and muons must be compatible with the hypothesis that they originate from the primary vertex and are required to satisfy the ``Gradient"~\cite{EGAM-2018-01} and ``PflowTight"~\cite{MUON-2018-03} isolation criteria, respectively.

Jets are reconstructed using the \antikt algorithm~\cite{Cacciari:2008gp, Fastjet}, with a radius parameter of \(R=0.4\), using particle-flow objects~\cite{PERF-2015-09} as input. Contamination from jets originating in additional $pp$ interactions per bunch crossing (pile-up) is reduced by using the jet-vertex-tagger algorithm~\cite{PERF-2014-03}.
At least two jets with $|\eta| < 4.5$ are required to be present in each event, with $\pt>65$~\GeV for the leading jet and $\pt>35~\GeV$ for the subleading jet.

The event selection in the signal region (SR) requires a same-sign lepton pair with an invariant mass $m_{\ell\ell}$ greater than $20$~\GeV. The missing transverse momentum, with magnitude \MET, must be larger than 30~\GeV in the SR to exploit the presence of neutrinos in the final state. The \MET is calculated from the negative vector sum of the transverse momenta of all of the selected and calibrated objects in the event including a track-based soft term~\cite{JETM-2020-03}. The VBS topology is characterized by at least two jets with a large invariant mass, $\mjj$, and a large absolute rapidity difference, $\left|\dyjj\right|$. The $\mjj$ of the two highest-$\pt$ jets must satisfy $\mjj>500~\GeV$ with $\left|\dyjj\right|>2.0$. Additional requirements to suppress background contributions from the Drell--Yan background in the dielectron final state, from top quark processes, and from processes with more than two leptons in the final state are described in Ref.~\cite{STDM-2018-32}. The data yield in the SR is 475 events. The expected purities in the SR of the EW and QCD \sswwjj processes and of their interference are approximately 58\%, 5.2\%, and, 1.6\%, respectively. The interference effects among the different \sswwjj polarization states in the SR are found to be negligible.

The SM production of $WZ/\gamma^*$ boson pairs in association with two jets, denoted as $\wzjj$, constitutes a large background, contributing to approximately 16\%\ of the overall expected event yield in the SR. It contributes when one of the leptons is not selected, typically because it is outside of the geometrical acceptance of the detector. The production of $\wzjj$ at LO has contributions both from processes that involve only electroweak (EW) interaction vertices (EW $\wzjj$) and from processes that involve strong interaction vertices (QCD $\wzjj$). The contributions of the EW and QCD $\wzjj$ processes are estimated with Monte Carlo (MC) simulated events. The modeling of the QCD $\wzjj$ process is constrained in a dedicated signal-depleted control region (CR) defined by requiring three charged leptons in the final state, $\mjj>200~\GeV$, and a trilepton invariant mass greater than 106~\GeV.

Leptons from hadron decays and jets misidentified as leptons are referred to as non-prompt leptons. The non-prompt lepton background arises mainly from $W$+jets and semileptonic $t\bar{t}$ processes. The non-prompt lepton, electron charge misidentification and photon conversion backgrounds are estimated using data-driven methods as described in Ref.~\cite{STDM-2018-32}, and contribute to approximately 18\% of the overall expected event yield. Events with the same selection criteria as in the SR but requiring $200 < \mjj < 500~\GeV$ are used to define the ``low-\mjj'' CR. This CR has a similar background composition as in the SR and is used to control the uncertainties of the major background contributions. Finally, small background contributions from triboson $VVV$ ($V=W/Z/\gamma$), $t\bar{t}V$, and $tZq$ processes are estimated using MC simulations and together contribute to less than 2\% of the expected event yield in the SR.

MC simulated event samples are used to model the signal $\sswwjj$ process with different polarization states, as well as other background processes. Simulated events are processed through the ATLAS simulation infrastructure~\cite{SOFT-2010-01} using \textsc{Geant4}~\cite{Agostinelli:2002hh}. The effect of pile-up is simulated by overlaying the hard-scattering process with minimum-bias events generated with \PYTHIA[8.186]~\cite{Sjostrand:2007gs} using the A3 set of tuned parameters~\cite{ATL-PHYS-PUB-2016-017}. Different pile-up conditions between data and simulation are taken into account by reweighting the mean number of interactions per bunch crossing in simulation to the number observed in data. Detailed descriptions of the simulated MC samples used for modeling the various background events can be found in Ref.~\cite{STDM-2018-32}. The large QCD $\wzjj$ background is simulated with the \SHERPA[2.2.12] event generator~\cite{Bothmann:2019yzt} using matrix elements that contain all diagrams with four EW vertices. This process is calculated for up to one additional parton at next-to-LO (NLO) and up to three additional partons at LO using \textsc{Comix}~\cite{Gleisberg:2008fv} and \OPENLOOPS~\cite{Cascioli:2011va}, and merged with the \SHERPA parton shower based on the Catani--Seymour dipole factorization~\cite{Schumann:2007mg}.

This analysis uses state-of-the-art calculations and simulations to incorporate the NLO EW and QCD effects in modeling the polarized EW \sswwjj production. The polarized on-shell \sswwll, \sswwlt, and \sswwtt signal processes are simulated in narrow-width approximation using \SHERPA[3]~\cite{Sherpa:2024mfk,Hoppe:2023uux} with up to one additional parton at LO in QCD and EW and merged with the \SHERPA parton shower. The \NNPDF[3.0nlo]~\cite{Ball:2014uwa} set of parton distribution function (PDF) is used. This multijet merging approach has been shown to capture the bulk of NLO QCD corrections to polarization fractions, as they typically stem from real emission rather than virtual diagrams~\cite{Hoppe:2023uux,Pelliccioli:2023zpd}, thus providing the most accurate hadron-level prediction available for these signal processes. Diagrams involving the $s$-channel and triboson contributions are omitted in this simulation by the use of the VBS approximation~\cite{Ballestrero:2018anz} to avoid possible divergences. The contributions of these missing diagrams and further missing off-shell production effects are estimated using unpolarized $\ell^\pm \nu \ell^\pm \nu jj$ \SHERPA[3] simulated samples at the same perturbative accuracy. These effects are taken into account by applying corrections to the signal predictions using all kinematic information in the full phase space via neural networks at particle level following the approach of Ref.~\cite{Andreassen:2019nnm}. The size of the corrections in the SR is approximately 10\%. In an alternative approach, the corrections are applied via particle level $\mjj$ reweighting and the resulting differences in the shapes of the signal distributions are considered as an uncertainty. This uncertainty is treated as uncorrelated among the \sswwll, \sswwlt, and \sswwtt signal processes to account for any residual differences in the size of the corrections.

The first calculations of the fixed-order next-to-LO (NLO) EW corrections for the production of polarized EW \sswwjj production have been recently performed~\cite{Denner:2024tlu}. The pure EW corrections are dominant and reduce the LO cross section of the EW \sswwjj process by approximately 15\% in the SR and have even larger impact in high-energy tails of distributions. The NLO EW corrections have different impacts for the \sswwll, \sswwlt, and \sswwtt processes with smaller in magnitude corrections for the longitudinal modes. Their effect is taken into account by applying $\mjj$ dependent factors to the signal samples. In an alternative approach, the corrections are applied as leading jet \pt dependent factors and the difference is considered as a systematic uncertainty.

Systematic uncertainties in this measurement arise from experimental and theory sources. The results are driven by the statistical uncertainty of the data in the SR and none of the considered systematic uncertainties have a significant impact on the sensitivity of this result. Experimental systematic uncertainties are related to the trigger, lepton reconstruction, identification and isolation efficiencies~\cite{MUON-2018-03, PERF-2017-01}, lepton energy (momentum) scale and resolution~\cite{MUON-2018-03, EGAM-2018-01}, jet energy scale and resolution~\cite{JETM-2018-05}, $b$-jet identification~\cite{FTAG-2018-01}, modeling of \MET~\cite{JETM-2020-03}, pile-up, and integrated luminosity~\cite{DAPR-2021-01,LUCID2}. The uncertainties related to the non-prompt lepton, electron charge misidentification, and photon conversion backgrounds are described in Ref.~\cite{STDM-2018-32}. Theoretical uncertainties associated with the signal and other background processes are evaluated using simulation. The theory uncertainties in the physics modeling of the polarized \sswwjj processes are estimated by varying the factorization, renormalization, resummation, and merging scales~\cite{Hoppe:2023uux}, the strong coupling constant $\alpha_S$, and the choice of the PDF~\cite{Butterworth:2015oua}. Additional uncertainties related to the propagation of the NLO EW corrections and the corrections due to the off-shell production and missing diagrams are considered as described above. The systematic uncertainties with the largest impact on the result are related to the finite number of data events used for data-driven background estimates, the theory uncertainties in the physics modeling of the polarized \sswwjj production, and uncertainties in the jet and \MET scale and resolution.

\begin{figure}[t!bp]
\centering
\subfigure[]{\includegraphics[width=0.32\textwidth]{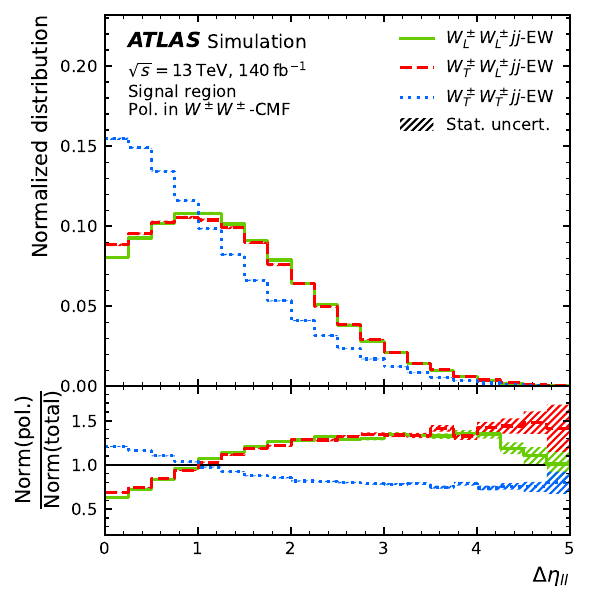}}
\subfigure[]{\includegraphics[width=0.32\textwidth]{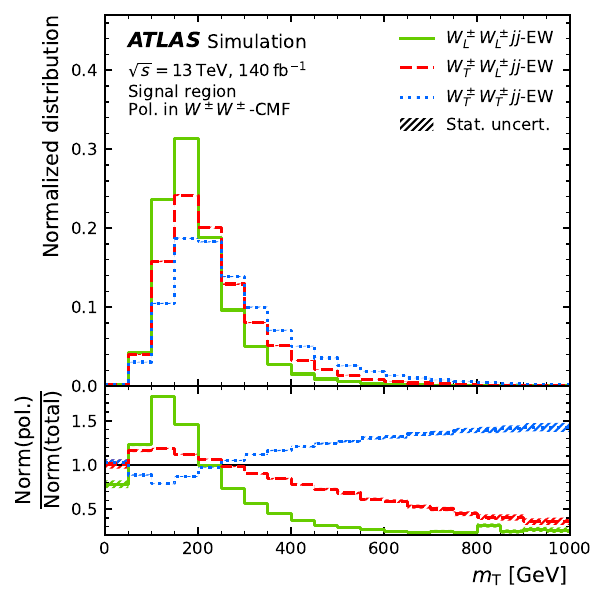}}
\subfigure[]{\includegraphics[width=0.32\textwidth]{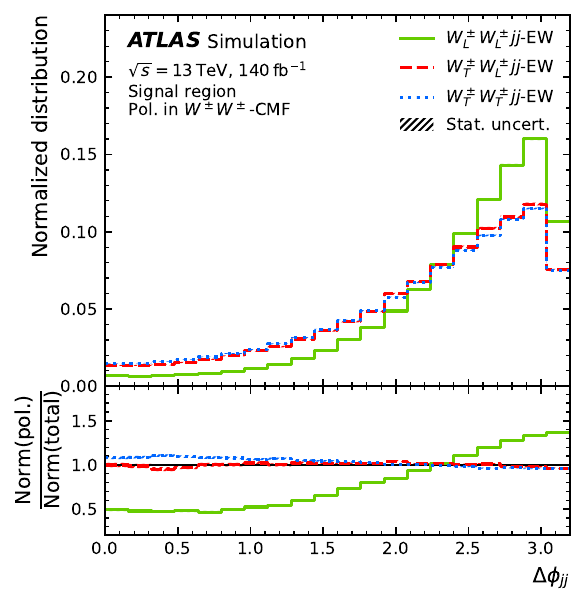}}
\subfigure[]{\includegraphics[width=0.32\textwidth]{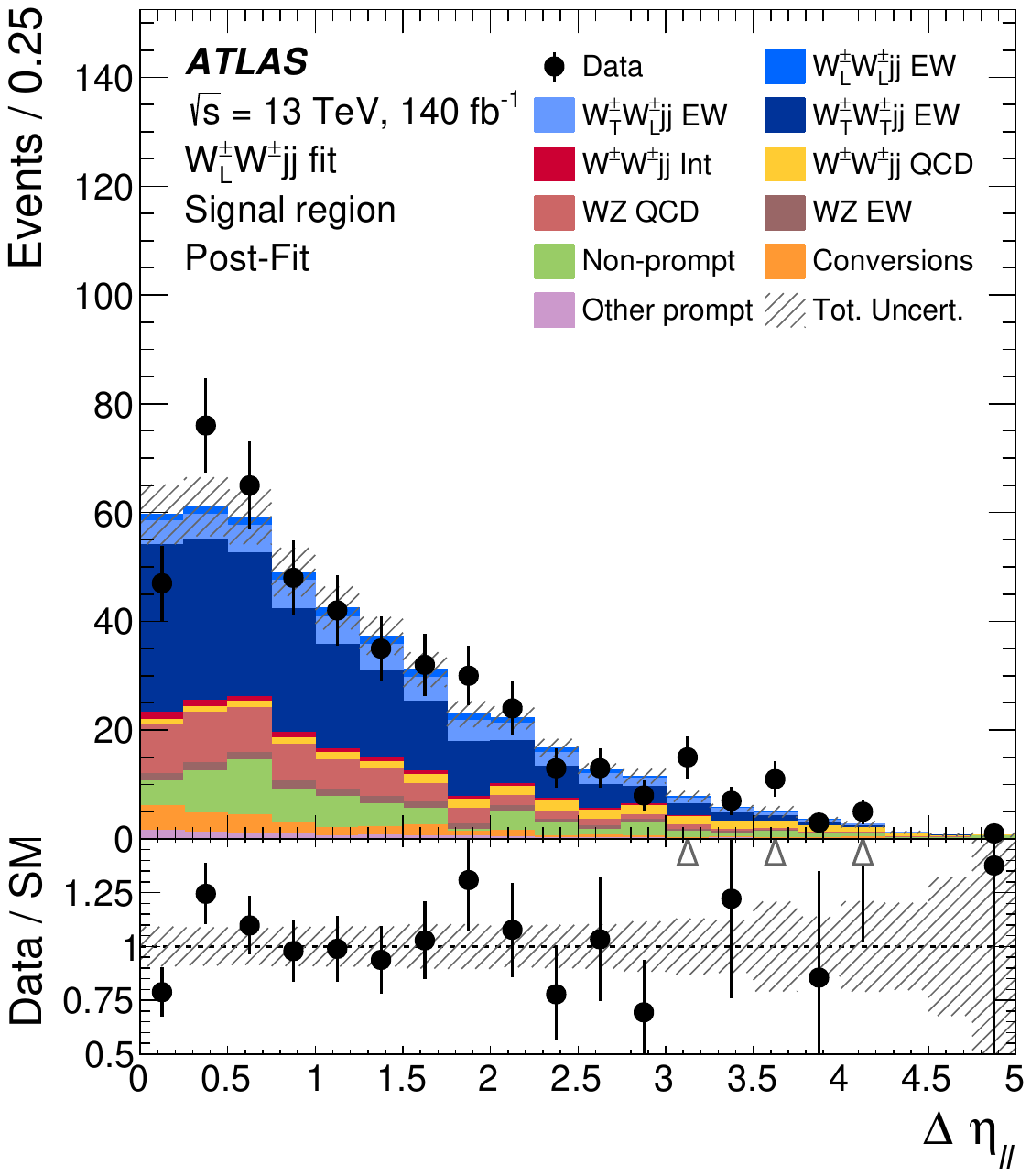}}
\subfigure[]{\includegraphics[width=0.32\textwidth]{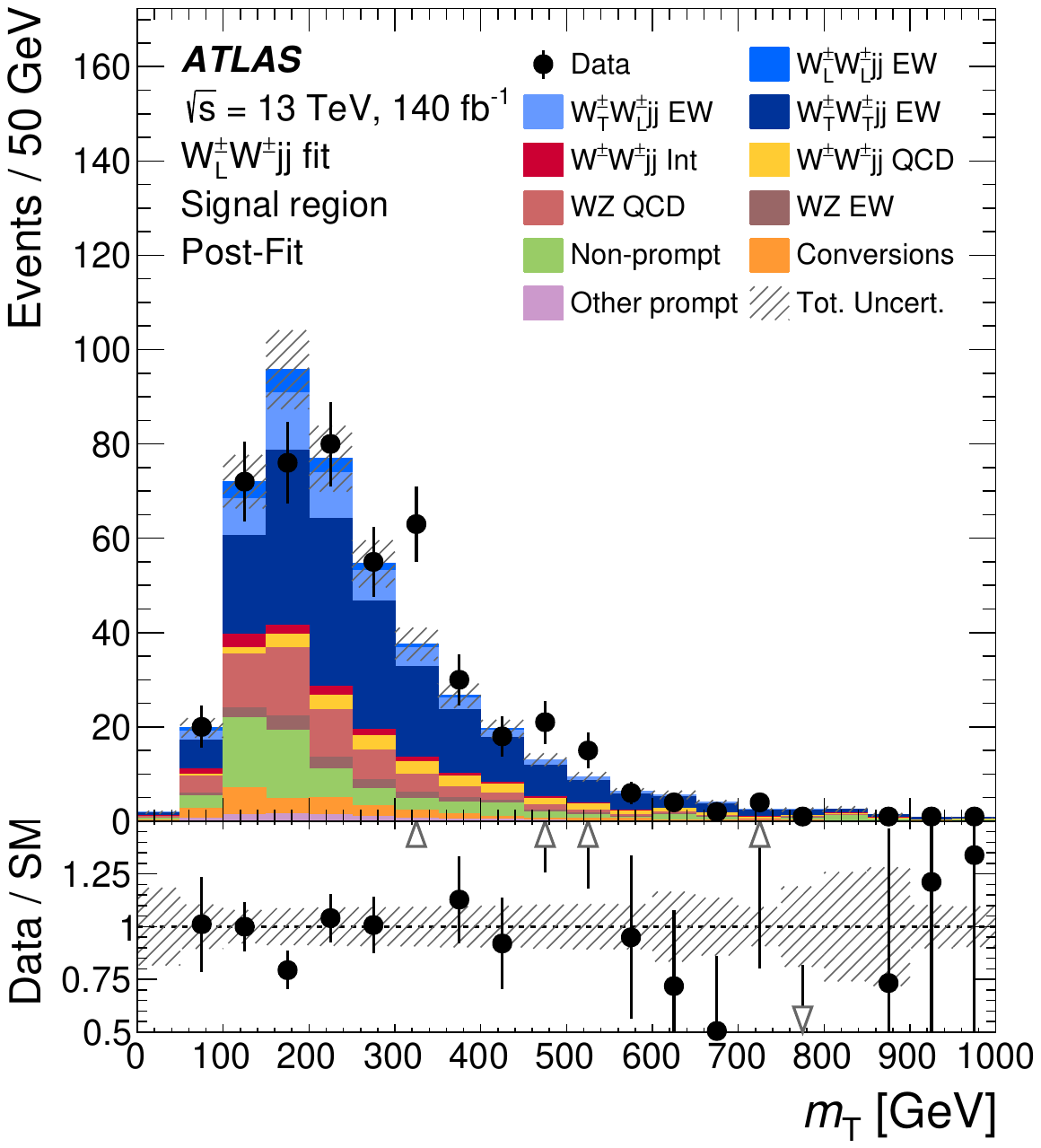}}
\subfigure[]{\includegraphics[width=0.32\textwidth]{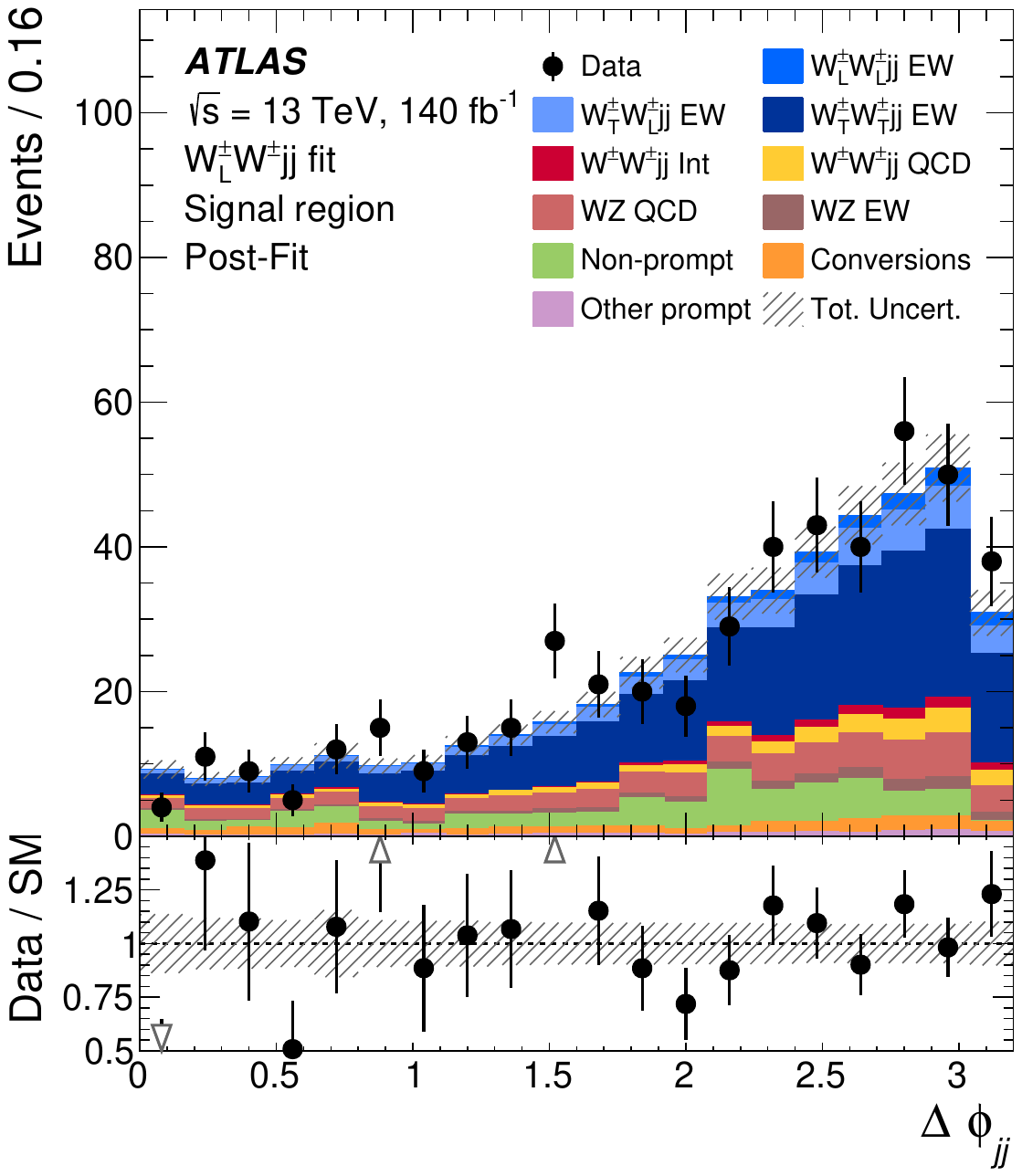}}
\caption{Distributions in the SR for the (a and d) difference in pseudorapidity between the leading and subleading leptons, $|\Delta \eta_{\ell\ell}|$, (b and e) transverse mass of the dilepton system and $\MET$~\cite{STDM-2018-32}, \mt, and (c and f)  difference in azimuthal angle between the leading and subleading jets, $\Delta \phi_{jj}$. The ratios of the contributions of each process to the total contribution are shown in the bottom panels in (a-c). The predicted yields are shown with their best-fit normalization and shape in (d-f). The shaded area surrounding the expectation represents the total uncertainties in the predicted yields. The ratios of the observed yields to the total SM predictions are shown by the points in the bottom panels. The ``Other prompt'' category combines $ZZ$, $VVV$, $t\bar{t}V$, and $tZq$ background processes. The polarization vectors are defined in the $\ssww$ center-of-mass reference frame.}
\label{fig:var_plots_main}
\end{figure}

To measure the production of the $\sswwll$ and $\sswwlx$ processes, a first DNN (inclusive DNN) is trained to separate the unpolarized EW \sswwjj production from the background processes. Two independent DNNs (signal DNNs) are then trained to extract the polarization information by separating the production of $\sswwll$ from $\sswwtx$ and $\sswwlx$ from $\sswwtt$. Dedicated optimizations of the model structures and hyperparameters are performed for each DNN with the open source framework OPTUNA~\cite{akiba2019optunanextgenerationhyperparameteroptimization}. The number of hidden layers are 9 and 10 for the inclusive and signal DNNs, respectively, with up to 204 neurons per layer used for the $\sswwlx$ signal DNN. The kinematic variables showing the best discrimination, evaluated by removing each variable and retraining the network, are retained.  Up to 20 such variables are retained for each DNN from a larger set. The variable importance is also evaluated with the \textsc{SHAP} framework~\cite{NIPS2017_8a20a862}. The \mjj variable provides a strong discrimination for the inclusive DNN as background processes tend to have smaller \mjj values than those from the EW \sswwjj production. Similarly, angular variables such as the differences in the azimuthal angle, pseudorapidity, and distance $\Delta R$ between the leading and subleading leptons and jets provide an important separation for the signal DNNs. Kinematic variables related to the transverse mass and $\pt$ of the systems constructed from the leptons, jets, and $\MET$ are also considered. The distributions of selected kinematic variables showing a strong discrimination for the $\sswwlx$ signal DNN are shown in Figure~\ref{fig:var_plots_main}. Events from $\sswwll$ generally exhibit larger |$\Delta \eta_{\ell\ell}$| and $\Delta \phi_{jj}$, and smaller $\mt$, than those from $\sswwtt$. A complete list of variables used for the DNNs, along with the distributions of additional discriminating variables, is discussed in the Appendix. The modeling of these DNN scores for the background processes is validated in the low-\mjj CR.

Binned maximum-likelihood fits are performed separately for the $\sswwll$ and $\sswwlx$ processes using the inclusive and the corresponding signal DNN score distributions. The signal DNN score distributions in three regions of the inclusive DNN score, with boundaries at (0, 0.2, 0.6, 1.0) and (0, 0.3, 0.7, 1.0), are fitted for the $\sswwll$ and $\sswwlx$ processes, respectively. The region boundaries are chosen independently to maximize the significances of the $\sswwll$ and $\sswwlx$ processes while ensuring that each bin contains at least 10 events originating from the other processes. For each fit, the three regions of the inclusive DNN score in the SR, the low-\mjj and QCD $\wzjj$ CRs are fitted simultaneously with the $\sswwll$ ($\sswwlx$), $\sswwtx$ ($\sswwtt$), and QCD $\wzjj$ normalizations kept as floating parameters. The relatively small contributions of the EW $\wzjj$ and QCD $\sswwjj$ processes are normalized to the SM predictions and allowed to vary within their uncertainties. The systematic uncertainties are included as nuisance parameters~\cite{Rolke:2004mj} with Gaussian priors. The nuisance parameters are adjusted in the fit with the shape and normalization of each distribution varying within the specified constraints.

\begin{figure}[t!bp]
\centering
\subfigure[]{\includegraphics[width=0.32\textwidth]{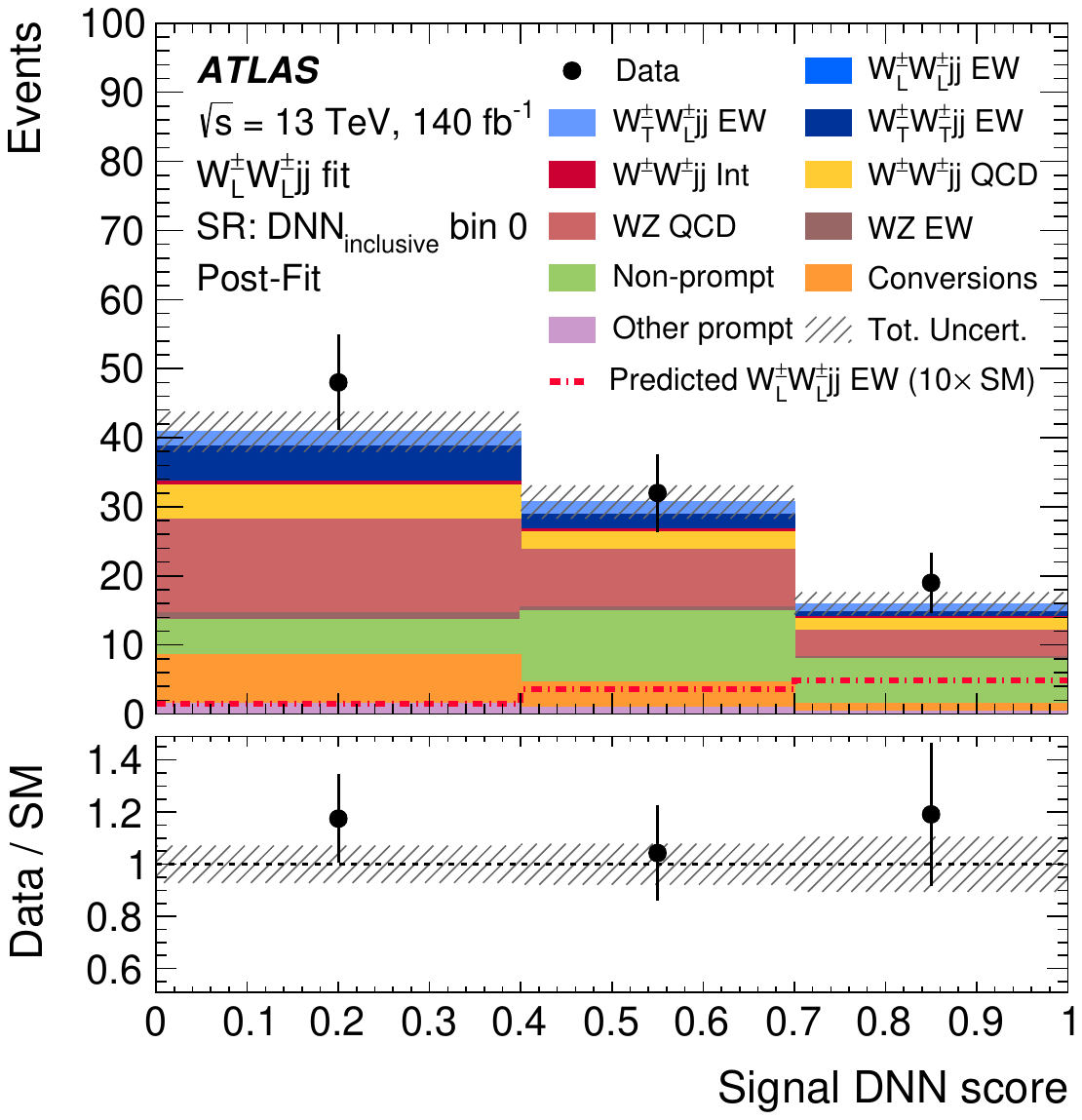}}
\subfigure[]{\includegraphics[width=0.32\textwidth]{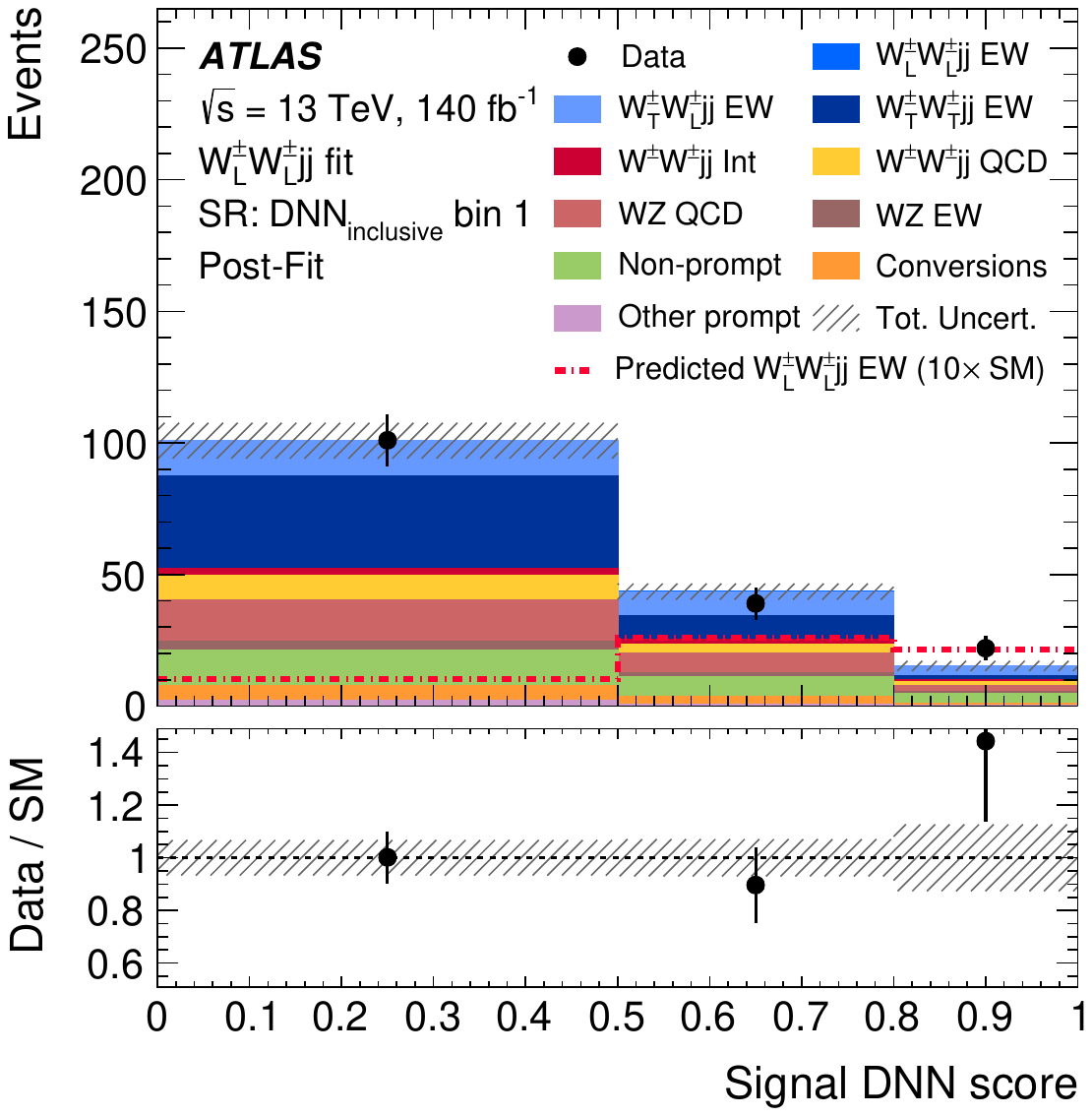}}
\subfigure[]{\includegraphics[width=0.32\textwidth]{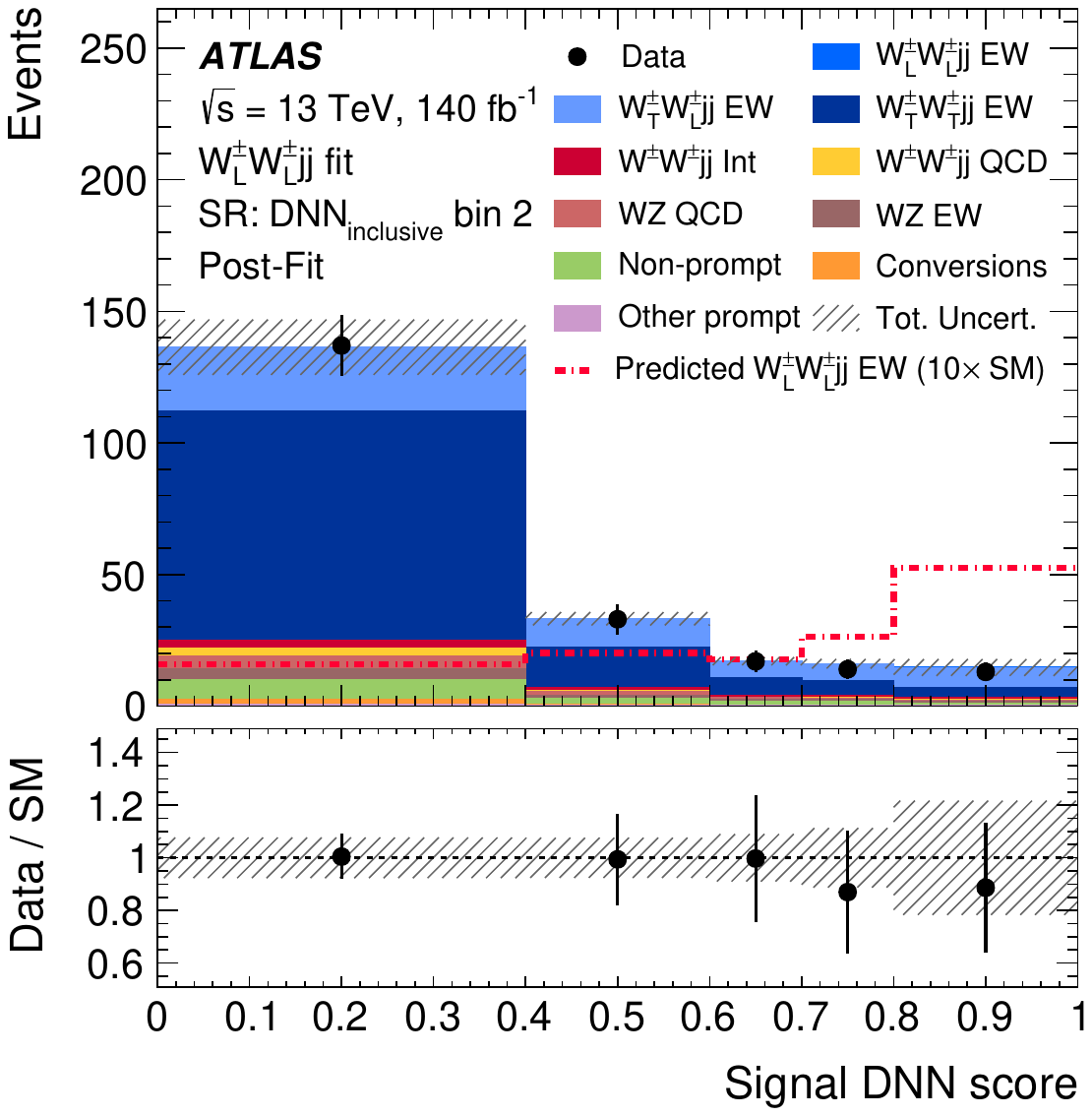}}
\subfigure[]{\includegraphics[width=0.32\textwidth]{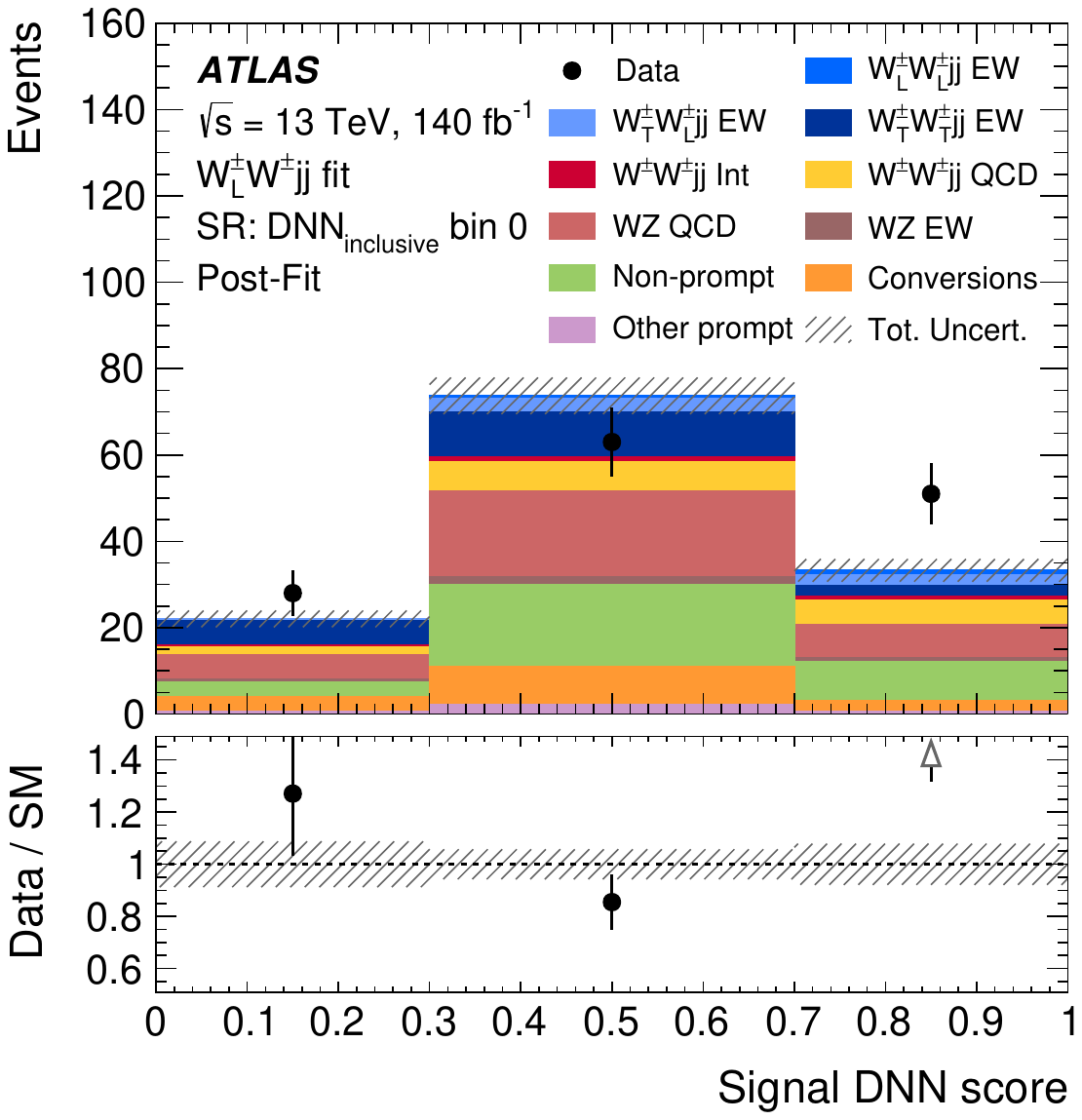}}
\subfigure[]{\includegraphics[width=0.32\textwidth]{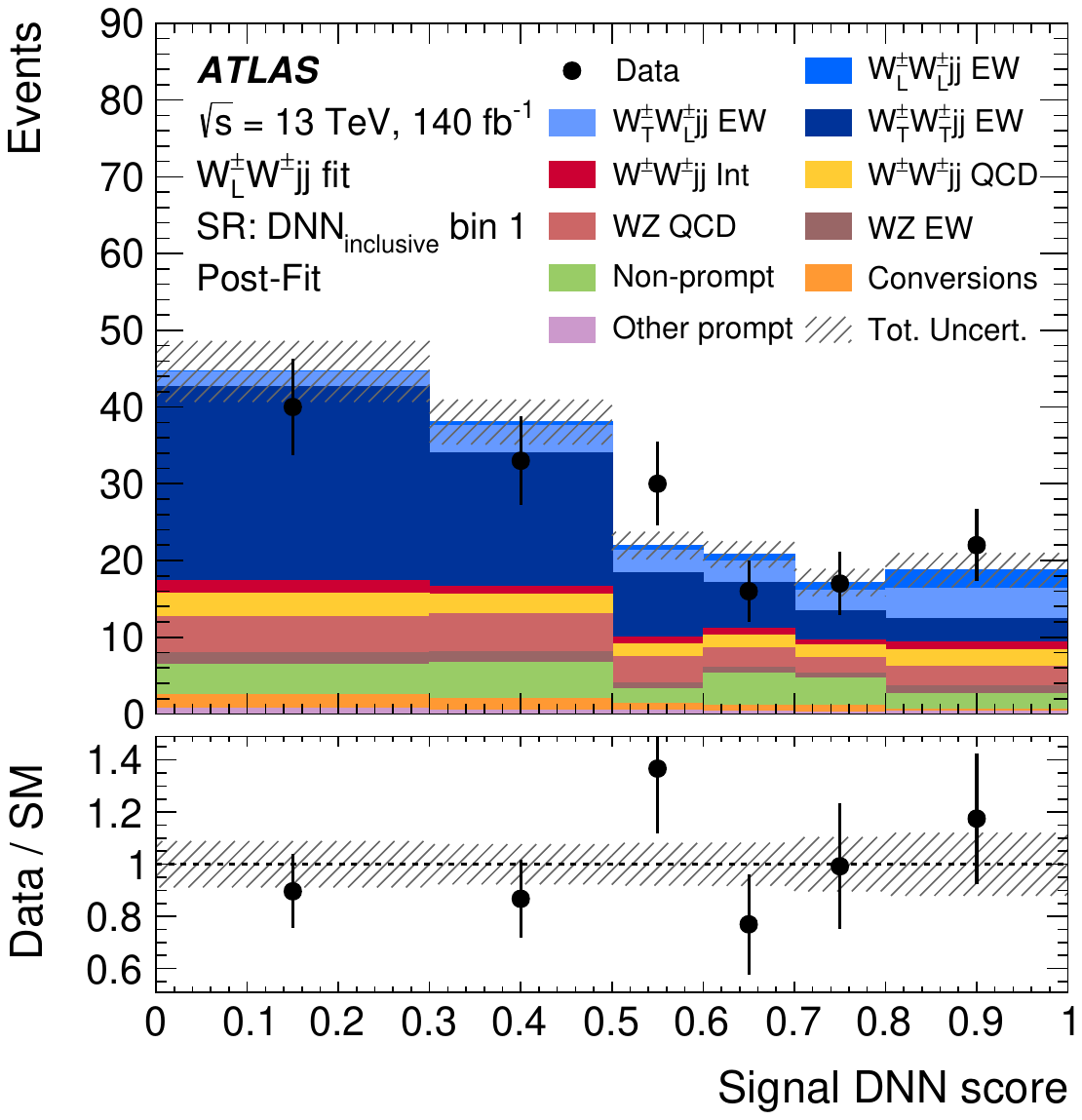}}
\subfigure[]{\includegraphics[width=0.32\textwidth]{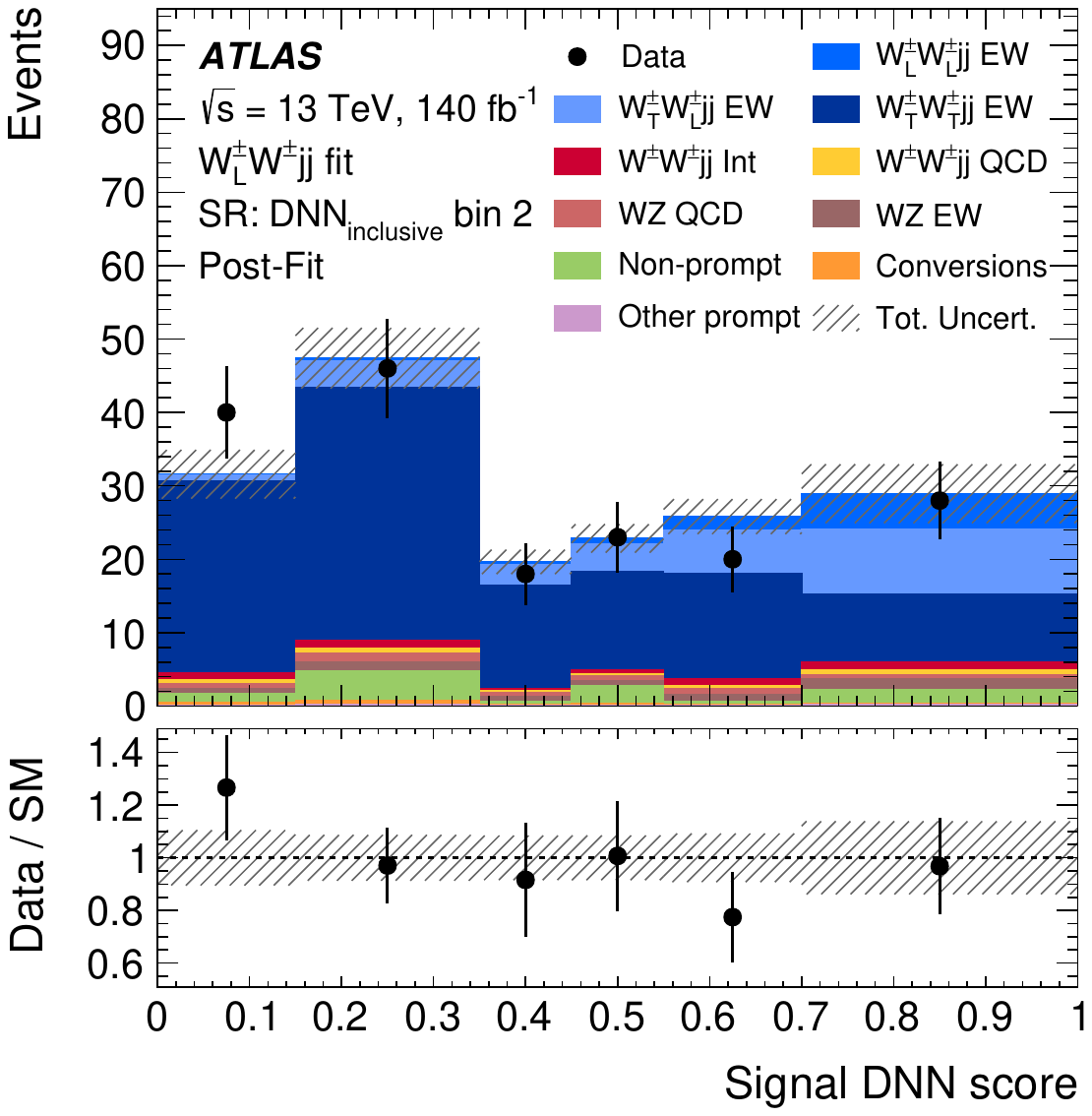}}
\caption{Distributions of the signal DNN score in the SR for the $\sswwll$ measurement in regions of the inclusive DNN with score values of (a) 0 to 0.2,  (b) 0.2 to 0.6, and (c) 0.6 to 1, and for the $\sswwlx$ measurement in regions of the inclusive DNN with score values of (d) 0 to 0.3,  (e) 0.3 to 0.7, and (f) 0.7 to 1. The polarization vectors are defined in the $\ssww$ center-of-mass reference frame. The predicted yields are shown with their best-fit normalization and shape. The shaded area surrounding the expectation represents the total uncertainties in the predicted yields. The ratios of the observed yields to the total SM predictions are shown by the points in the bottom panels. The ``Other prompt'' category combines $ZZ$, $VVV$, $t\bar{t}V$, and $tZq$ background processes. The dashed lines in (a-c) show the SM predictions of the  $\sswwll$ contributions multiplied by a factor of 10.}
\label{fig:sr_plots}
\end{figure}

The signal DNN score distributions in each region of the inclusive DNN are shown in Figure~\ref{fig:sr_plots} with the normalizations and nuisance parameters adjusted by the fit. A good separation between the different polarization states is achieved.  The post-fit yields are shown in the Appendix. The normalization factors for the QCD $\wzjj$ process are $0.74\pm0.08$ and $0.75\pm0.08$ for the $\sswwll$ and $\sswwlx$ fits, respectively, consistent with the value reported in Ref.~\cite{STDM-2018-32}. No uncertainties are significantly constrained or pulled in the simultaneous fit.

The fiducial cross sections times branching fraction for the $\sswwll$ ($\sigma \mathcal{B} (\sswwll)$) and $\sswwlx$ ($\sigma \mathcal{B} (\sswwlx$)) processes are extracted in the fiducial region defined to be as close as possible to the detector region as described in Ref.~\cite{STDM-2018-32}. The fiducial phase space definition requires two prompt leptons at particle level ($e$ or $\mu$) with $\pT > 27~\GeV$ and $|\eta| < 2.5$, dressed by adding the four-momenta of nearby prompt photons within a small cone of $\Delta R<0.1$, with $\mll > 20~\GeV$. Contributions of events with $\tau$-leptons from at least one of the $W$ boson decays, with the $\tau$-lepton decaying leptonically to an electron or a muon, are excluded from the fiducial region definition. At least two jets are required at particle level with $|\eta|<4.5$ and $\pT > 65~\GeV\ (35~\GeV)$ for the leading (sub-leading) jet, with the $\mjj$ of the two highest-\pT jets greater than 500~\GeV and $\left|\dyjj\right|>2$.

No significant excess of events consistent with the $\sswwll$ production is observed. The observed (expected) 95\% confidence level (CL) upper limit on $\sigma \mathcal{B} (\sswwll)$ is 0.45 (0.70) fb. The 95\% CL limits are derived using the $\text{CL}_{\textrm{s}}$ method~\cite{Junk:1999kv,Read:2002hq}. The expected limit is evaluated using an Asimov data set~\cite{Cowan:2010js} assuming the SM contributions of the $\sswwll$ and $\sswwtx$ processes. Assuming the null hypothesis of no contribution from the $\sswwll$  process, an expected 95\% CL upper limit of 0.43 fb on $\sigma \mathcal{B} (\sswwll)$ is obtained using an Asimov data set. The significance of a non-zero $\sswwlx$ event yield is quantified assuming contributions only from the $\sswwtt$ process. The observed significance for the $\sswwlx$ process is 3.3 $\sigma$ with an expected significance, evaluated with an Asimov data set, of 4.0 $\sigma$. The asymptotic approximation~\cite{Cowan:2010js}, whose validity was confirmed through studies with pseudo-experiments, is used to calculate the limits and significances.

\begin{table}[t!bp]
\centering
\caption{Measured and predicted fiducial cross sections for the $\sswwll$ and the $\sswwtx$, and the $\sswwlx$ and the $\sswwtt$ processes. The polarization vectors are defined in the $\ssww$ center-of-mass reference frame. The total measurement uncertainty is shown along with the statistical, modeling systematic and experimental systematic uncertainties. The modeling uncertainties include the theory uncertainties affecting the signal and background processes as well as the effect of a finite number of MC events and of data events used for data-driven background estimates. The predicted cross sections from \SHERPA[3] include the NLO EW corrections and the corrections to take into account the off-shell production and missing contributions. The predicted cross section uncertainties include the PDF, $\alpha_S$, parton shower, and scale variations.}
\small
\begin{tabular}{cccc}
\toprule
Process & Predicted $\sigma \mathcal{B}$ (fb) & Measured $\sigma \mathcal{B}$ (fb) & Uncertainty breakdown (fb) \\
\midrule
\sswwll & \ensuremath{0.29 \pm 0.07} & \ensuremath{0.01 \pm 0.21 \text{\xspace(tot.)}} & \ensuremath{ \pm 0.20 \text{\xspace(stat.)} \pm 0.02 \text{\xspace(mod. syst.)} \pm 0.05 \text{\xspace(exp. syst.)}} \\
\sswwtx & \ensuremath{2.56 \pm 0.64}  & \ensuremath{3.39 \pm 0.35 \text{\xspace(tot.)}} & \ensuremath{ \pm 0.30 \text{\xspace(stat.)} \pm 0.08 \text{\xspace(mod. syst.)} \pm 0.16 \text{\xspace(exp. syst.)}} \\
\hline
\sswwlx & \ensuremath{1.18 \pm 0.29} & \ensuremath{0.88 \pm 0.30 \text{\xspace(tot.)}} & \ensuremath{ \pm 0.28 \text{\xspace(stat.)} \pm 0.05 \text{\xspace(mod. syst.)} \pm 0.08 \text{\xspace(exp. syst.)}} \\
\sswwtt & \ensuremath{1.67 \pm 0.40} & \ensuremath{2.49 \pm 0.32 \text{\xspace(tot.)}} & \ensuremath{ \pm 0.30 \text{\xspace(stat.)} \pm 0.05 \text{\xspace(mod. syst.)} \pm 0.12 \text{\xspace(exp. syst.)}} \\
\bottomrule
\end{tabular}
\label{tab:fiducial}
\vspace{10pt}
\end{table}

The measured $\sigma \mathcal{B} (\sswwll)$ and $\sigma \mathcal{B} (\sswwtx)$, and $\sigma \mathcal{B} (\sswwlx)$ and $\sigma \mathcal{B} (\sswwtt$), and the corresponding theoretical predictions from \SHERPA[3] are summarized in Table~\ref{tab:fiducial}. The NLO EW corrections and the corrections to take into account the off-shell production and missing contributions as described earlier are included in the predictions. The uncertainties in the predictions are dominated by the scale variations, with an uncertainty of about 24\%. The measured cross sections are in agreement with the SM predictions. The results of the studies performed using the initial-state parton--parton reference frame to define the polarization vectors are shown in the Appendix.

In summary, this Letter reports the first evidence of EW production of polarized same-sign $W$ boson pairs where at least one of the $W$ bosons is longitudinally polarized and the most stringent limits to date on the fiducial cross section of two longitudinally polarized same-sign $W$ bosons. The data set used corresponds to an integrated luminosity of 140~fb$^{-1}$ of proton--proton collisions at a center-of-mass energy of 13~\TeV, collected with the ATLAS detector at the LHC. The study is performed in final states including two same-sign leptons (electrons or muons), missing transverse momentum, and at least two jets with large invariant mass and a large rapidity difference. An observed (expected) 95\% CL upper limit of 0.45 (0.70) fb is reported on the EW production of two longitudinally polarized same-sign $W$ bosons. The observed significance of the production with at least one longitudinally polarized $W$ boson is 3.3 standard deviations with a measured fiducial cross section of $0.88 \pm 0.30$ fb, in agreement with the SM prediction.

\section*{Acknowledgements}
%

%
%

%
%

We thank CERN for the very successful operation of the LHC and its injectors, as well as the support staff at
CERN and at our institutions worldwide without whom ATLAS could not be operated efficiently.

The crucial computing support from all WLCG partners is acknowledged gratefully, in particular from CERN, the ATLAS Tier-1 facilities at TRIUMF/SFU (Canada), NDGF (Denmark, Norway, Sweden), CC-IN2P3 (France), KIT/GridKA (Germany), INFN-CNAF (Italy), NL-T1 (Netherlands), PIC (Spain), RAL (UK) and BNL (USA), the Tier-2 facilities worldwide and large non-WLCG resource providers. Major contributors of computing resources are listed in Ref.~\cite{ATL-SOFT-PUB-2025-001}.

We gratefully acknowledge the support of ANPCyT, Argentina; YerPhI, Armenia; ARC, Australia; BMWFW and FWF, Austria; ANAS, Azerbaijan; CNPq and FAPESP, Brazil; NSERC, NRC and CFI, Canada; CERN; ANID, Chile; CAS, MOST and NSFC, China; Minciencias, Colombia; MEYS CR, Czech Republic; DNRF and DNSRC, Denmark; IN2P3-CNRS and CEA-DRF/IRFU, France; SRNSFG, Georgia; BMFTR, HGF and MPG, Germany; GSRI, Greece; RGC and Hong Kong SAR, China; ICHEP and Academy of Sciences and Humanities, Israel; INFN, Italy; MEXT and JSPS, Japan; CNRST, Morocco; NWO, Netherlands; RCN, Norway; MNiSW, Poland; FCT, Portugal; MNE/IFA, Romania; MSTDI, Serbia; MSSR, Slovakia; ARIS and MVZI, Slovenia; DSI/NRF, South Africa; MICIU/AEI, Spain; SRC and Wallenberg Foundation, Sweden; SERI, SNSF and Cantons of Bern and Geneva, Switzerland; NSTC, Taipei; TENMAK, T\"urkiye; STFC/UKRI, United Kingdom; DOE and NSF, United States of America.

Individual groups and members have received support from BCKDF, CANARIE, CRC and DRAC, Canada; CERN-CZ, FORTE and PRIMUS, Czech Republic; COST, ERC, ERDF, Horizon 2020, ICSC-NextGenerationEU and Marie Sk{\l}odowska-Curie Actions, European Union; Investissements d'Avenir Labex, Investissements d'Avenir Idex and ANR, France; DFG and AvH Foundation, Germany; Herakleitos, Thales and Aristeia programmes co-financed by EU-ESF and the Greek NSRF, Greece; BSF-NSF and MINERVA, Israel; NCN and NAWA, Poland; La Caixa Banking Foundation, CERCA Programme Generalitat de Catalunya and PROMETEO and GenT Programmes Generalitat Valenciana, Spain; G\"{o}ran Gustafssons Stiftelse, Sweden; The Royal Society and Leverhulme Trust, United Kingdom.

In addition, individual members wish to acknowledge support from Armenia: Yerevan Physics Institute (FAPERJ); CERN: European Organization for Nuclear Research (CERN DOCT); Chile: Agencia Nacional de Investigaci\'on y Desarrollo (FONDECYT 1230812, FONDECYT 1240864); China: Chinese Ministry of Science and Technology (MOST-2023YFA1605700, MOST-2023YFA1609300), National Natural Science Foundation of China (NSFC - 12175119, NSFC 12275265); Czech Republic: Czech Science Foundation (GACR - 24-11373S), Ministry of Education Youth and Sports (ERC-CZ-LL2327, FORTE CZ.02.01.01/00/22\_008/0004632), PRIMUS Research Programme (PRIMUS/21/SCI/017); EU: H2020 European Research Council (ERC - 101002463); European Union: European Research Council (BARD No. 101116429, ERC - 948254, ERC 101089007), European Regional Development Fund (SMASH COFUND 101081355, SLO ERDF), Horizon 2020 Framework Programme (MUCCA - CHIST-ERA-19-XAI-00), European Union, Future Artificial Intelligence Research (FAIR-NextGenerationEU PE00000013), Horizon 2020 (EuroHPC - EHPC-DEV-2024D11-051), Italian Center for High Performance Computing, Big Data and Quantum Computing (ICSC, NextGenerationEU); France: Agence Nationale de la Recherche (ANR-21-CE31-0022, ANR-22-EDIR-0002); Germany: Baden-Württemberg Stiftung (BW Stiftung-Postdoc Eliteprogramme), Deutsche Forschungsgemeinschaft (DFG - 469666862, DFG - CR 312/5-2); China: Research Grants Council (GRF); Italy: Istituto Nazionale di Fisica Nucleare (ICSC, NextGenerationEU), Ministero dell'Università e della Ricerca (NextGenEU I53D23000820006 M4C2.1.1); Japan: Japan Society for the Promotion of Science (JSPS KAKENHI JP22H01227, JSPS KAKENHI JP22H04944, JSPS KAKENHI JP22KK0227, JSPS KAKENHI JP23KK0245, JSPS KAKENHI JP24K23939); Norway: Research Council of Norway (RCN-314472); Poland: Ministry of Science and Higher Education (IDUB AGH, POB8, D4 no 9722), Polish National Science Centre (NCN 2021/42/E/ST2/00350, NCN OPUS 2023/51/B/ST2/02507, NCN OPUS nr 2022/47/B/ST2/03059, NCN UMO-2019/34/E/ST2/00393, UMO-2022/47/O/ST2/00148, UMO-2023/49/B/ST2/04085, UMO-2023/51/B/ST2/00920, UMO-2024/53/N/ST2/00869); Portugal: Foundation for Science and Technology (FCT); Spain: Generalitat Valenciana (Artemisa, FEDER, IDIFEDER/2018/048), Ministry of Science and Innovation (MCIN \& NextGenEU PCI2022-135018-2, MICIN \& FEDER PID2021-125273NB, RYC2019-028510-I, RYC2020-030254-I, RYC2021-031273-I, RYC2022-038164-I); Sweden: Carl Trygger Foundation (Carl Trygger Foundation CTS 22:2312), Swedish Research Council (Swedish Research Council 2023-04654, VR 2021-03651, VR 2022-03845, VR 2022-04683, VR 2023-03403, VR 2024-05451), Knut and Alice Wallenberg Foundation (KAW 2018.0458, KAW 2022.0358, KAW 2023.0366); Switzerland: Swiss National Science Foundation (SNSF - PCEFP2\_194658); United Kingdom: Leverhulme Trust (Leverhulme Trust RPG-2020-004), Royal Society (NIF-R1-231091); United States of America: U.S. Department of Energy (ECA DE-AC02-76SF00515), Neubauer Family Foundation.

%
%


%
\clearpage
\appendix
\part*{Appendix}
\addcontentsline{toc}{part}{Appendix}

\begin{figure}[t!bp]
\centering
\subfigure[]{\includegraphics[width=0.32\textwidth]{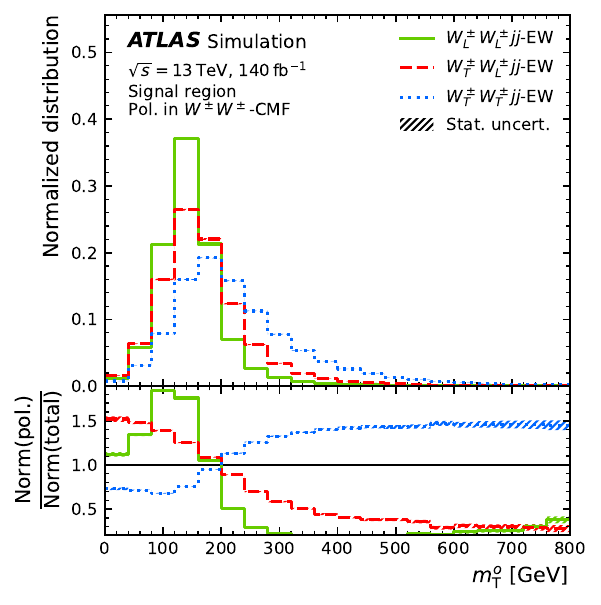}}
\subfigure[]{\includegraphics[width=0.32\textwidth]{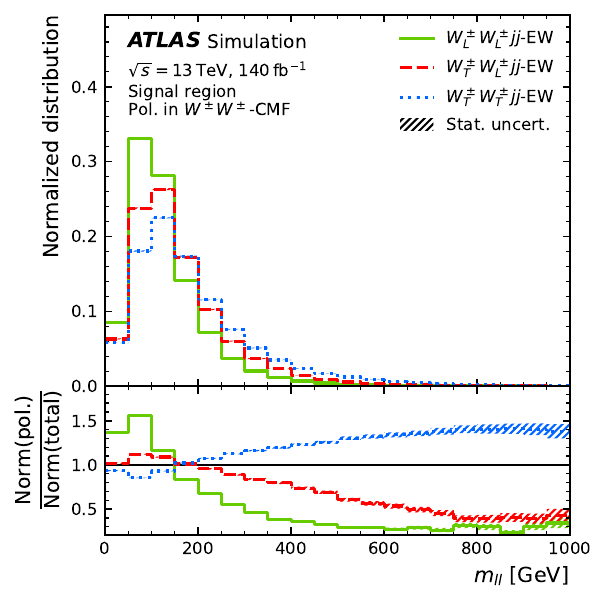}}
\subfigure[]{\includegraphics[width=0.32\textwidth]{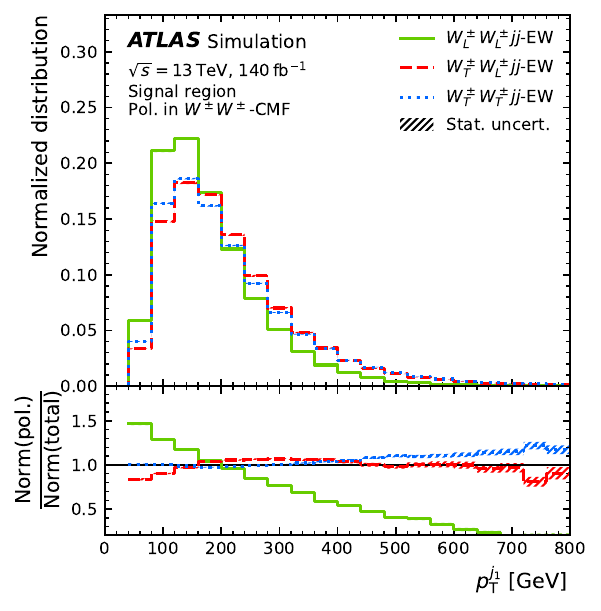}}
\subfigure[]{\includegraphics[width=0.32\textwidth]{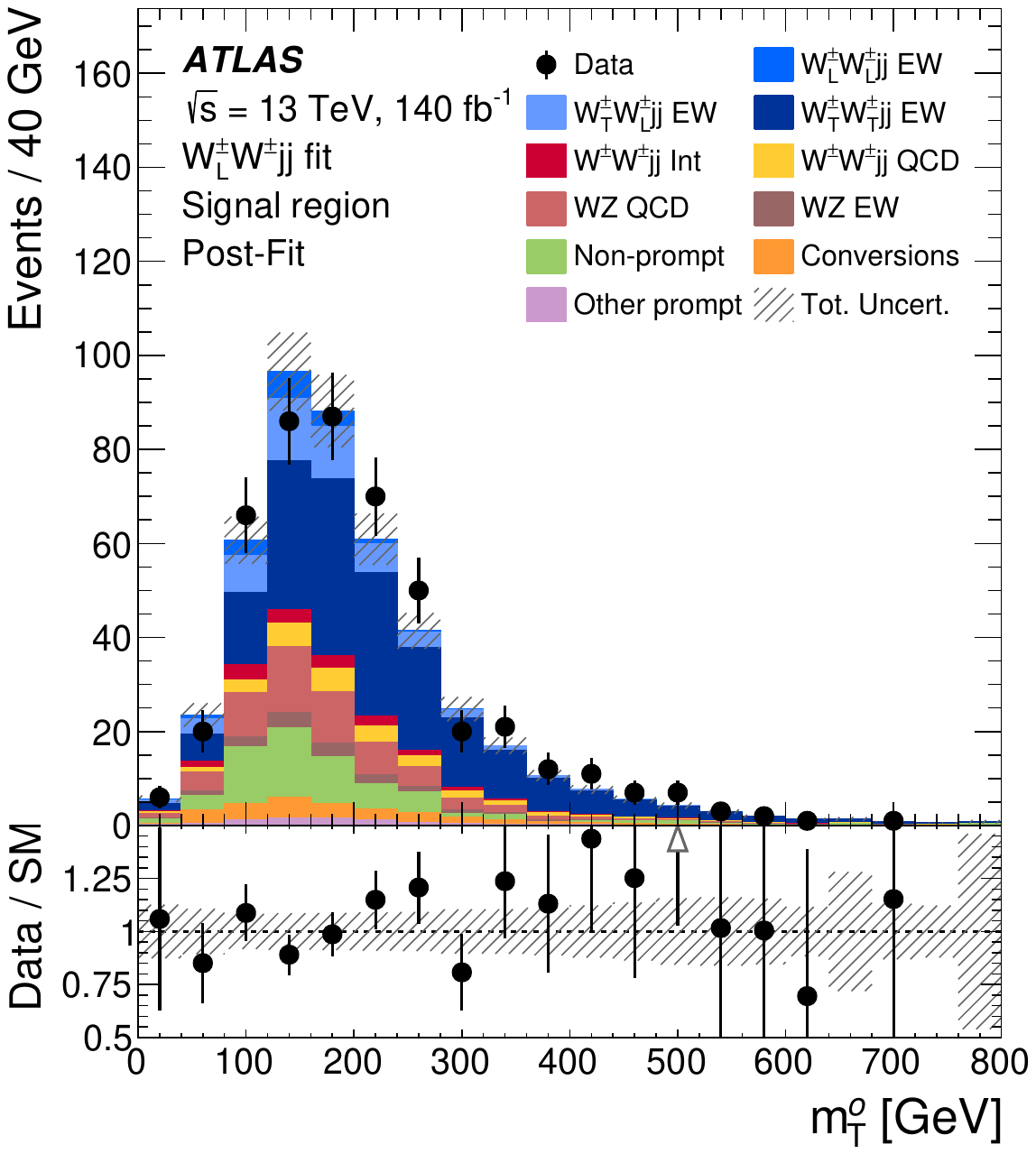}}
\subfigure[]{\includegraphics[width=0.32\textwidth]{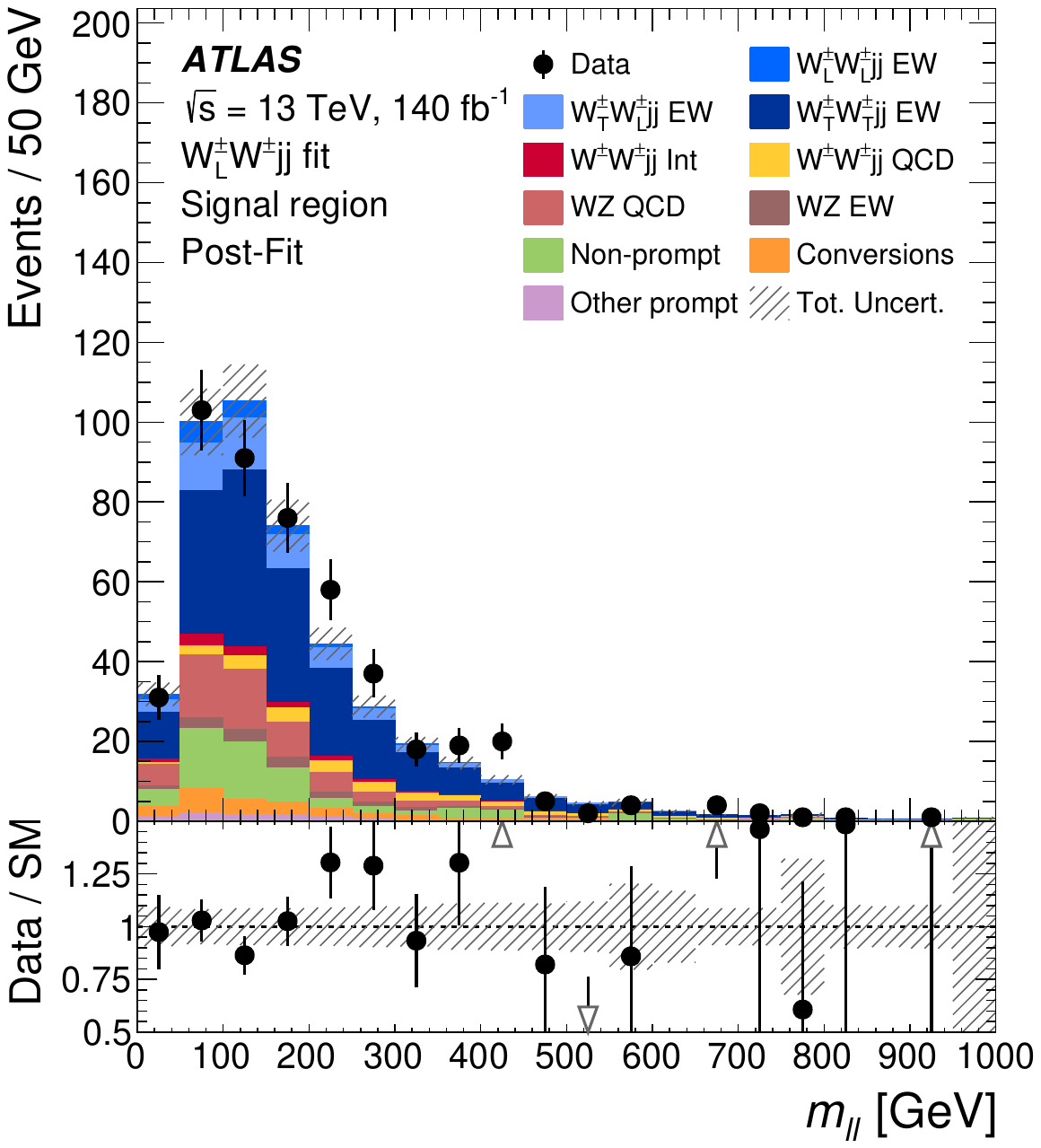}}
\subfigure[]{\includegraphics[width=0.32\textwidth]{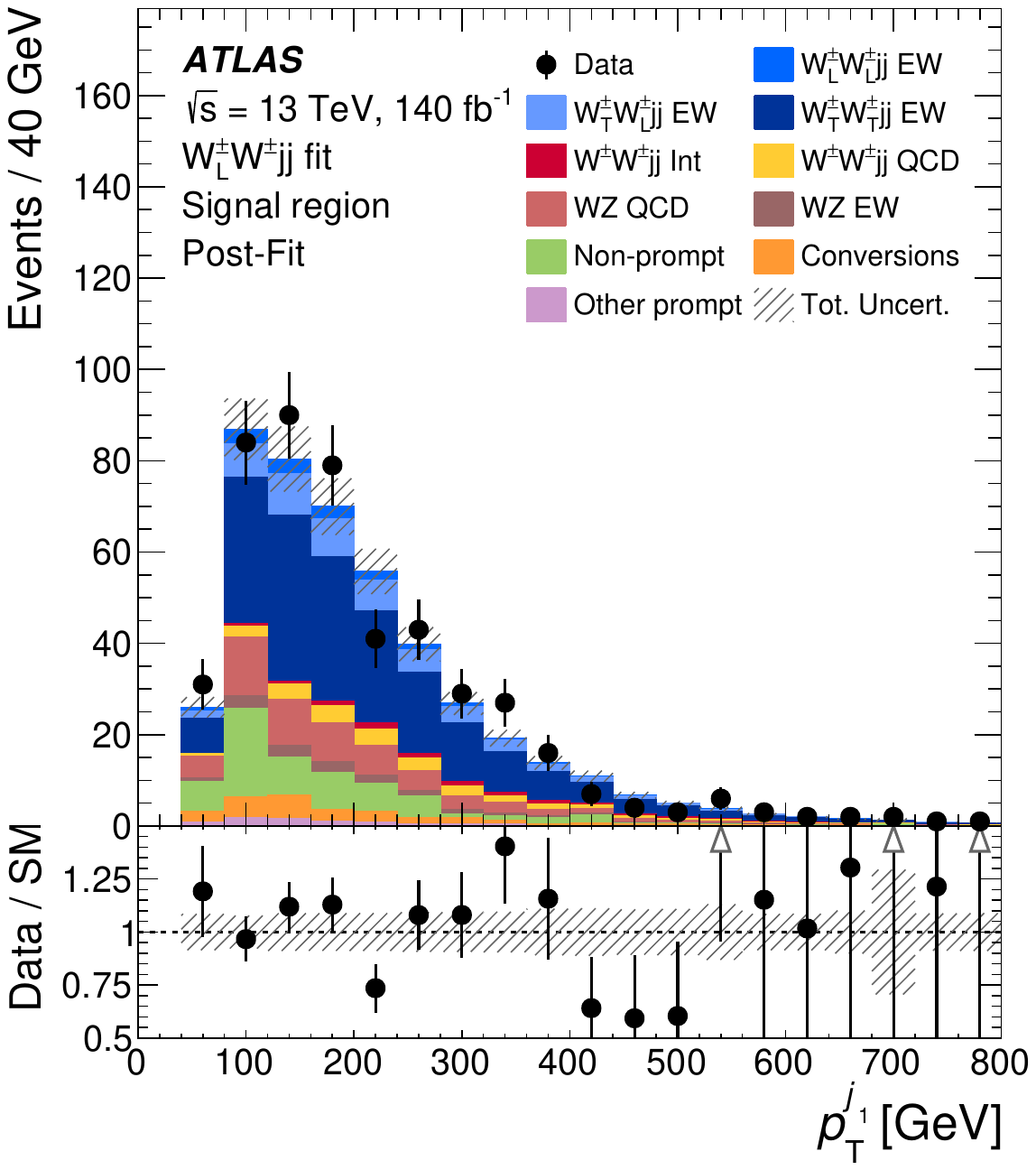}}
\caption{Distributions in the SR for the (a and d) $\mt^{o}$, (b and e)  $m_{\ell\ell}$, and (c and f) $\pt^{j_1}$. The ratios of the contributions of each process to the total contribution are shown in the bottom panels in (a-c). The predicted yields are shown with their best-fit normalization and shape in (d-f). The shaded area surrounding the expectation represents the total uncertainties in the predicted yields. The ratios of the observed yields to the total SM predictions are shown by the points in the bottom panels. The ``Other prompt'' category combines $ZZ$, $VVV$, $t\bar{t}V$, and $tZq$ background processes. The polarization vectors are defined in the $\ssww$ center-of-mass reference frame.}
\label{fig:var_plots_app}
\end{figure}

The variables used in the training and optimization of the DNNs are summarized in Table~\ref{table:MVA_training_variables}. The distributions of selected variables showing a significant separation between the \sswwll, \sswwlt, and \sswwtt processes are shown in Figure~\ref{fig:var_plots_app}. The expected signal and background yields in the SR for different regions of the inclusive DNN score are shown in Table~\ref{tab:yields}. A summary of the fractional uncertainties on $\sigma \mathcal{B} (\sswwlx$) measurement is shown in Table~\ref{tab:unc}.

The results of the studies performed using the initial-state parton--parton reference frame to define the polarization vectors are reported. The measured $\sigma \mathcal{B} (\sswwll)$ and $\sigma \mathcal{B} (\sswwtx)$, and $\sigma \mathcal{B} (\sswwlx)$ and $\sigma \mathcal{B} (\sswwtt$), and the corresponding theoretical predictions from \SHERPA[3] are shown in Table~\ref{tab:fiducialpp}. The observed significance for the $\sswwlx$ process is 2.5 $\sigma$ with an expected significance of 3.5 $\sigma$. The observed (expected) 95\% CL upper limit on $\sigma \mathcal{B} (\sswwll)$ is 0.60 (0.63) fb.  The expected limit is evaluated using an Asimov data set assuming the SM contributions of the $\sswwll$ and $\sswwtx$ processes. An expected limit of 0.46 fb is obtained using an Asimov data set assuming the null hypothesis of no contribution from the $\sswwll$ process.

\begin{table}[t!bp]
\centering
\caption{Variables used in the training and optimization of the DNNs. The variables are scaled with different scaling functions for the training and application.}
\resizebox{\linewidth}{!}{
\begin{tabular}{|l|c|c|}
\hline
Kinematics &  Descriptions & Scaling \\
\hline
\multicolumn{3}{|c|}{Object-level variables} \\
\hline
$\pt^{\ell_1}$ & $p_T$ of the leading lepton & $\log_{10}(x)$ \\
$\eta^{\ell1}$ & $\eta$ of the leading lepton & $x$ \\
$\ell_1$ type & Flavor of the leading lepton & $x$ \\
$\pt^{\ell_2}$ & $\pt$ of the subleading lepton & $\log_{10}(x)$ \\
$\eta^{\ell_2}$ & $\eta$ of the subleading lepton & $x$ \\
$\phi^{\ell_2}-\phi^{\ell_1}$ & Difference in azimuthal angle between the leading and subleading leptons & $x$ \\
$\ell_2$ type & Flavor of the subleading lepton & $x$ \\
$\pt^{j_1}$ & $\pt$ of the leading jet & $\log_{10}(x)$ \\
$\eta^{j_1}$ & $\eta$ of the leading jet & $x$ \\
$\phi^{j_1}-\phi^{\ell_1}$ &  Difference in azimuthal angle between the leading jet and leading lepton & $x$ \\
$\pt^{j_2}$ & $\pt$ of the subleading jet & $\log_{10}(x)$ \\
$\eta^{j_2}$ & $\eta$ of the subleading jet & $x$ \\
$\phi^{j_2}-\phi^{\ell_1}$ & Difference in azimuthal angle between the subleading jet and leading lepton  & $x$ \\
$\MET$ & Missing transverse momentum & $\log_{10}(x)$ \\
$\phi(\MET)-\phi^{\ell_1}$ & Difference in azimuthal angle between the missing transverse momentum and leading lepton & $x$ \\
\hline
\multicolumn{3}{|c|}{Event-level variables} \\
\hline
$Z^{\star}_{\ell_1}$ & Zeppenfeld variable of the leading lepton~\cite{Rainwater:1996ud} & $\sqrt{x}$ \\
$Z^{\star}_{\ell_2}$	 &	Zeppenfeld variable of the subleading lepton~\cite{Rainwater:1996ud} & $\sqrt{x}$ \\
$\mt^{\ell_1}$ & Transverse mass of the leading lepton and $\MET$ & $\sqrt{x}$ \\
$\mt^{\ell_2}$ & Transverse mass of the subleading lepton and $\MET$ & $\sqrt{x}$ \\
$\Delta R_{\ell\ell}$ & Distance $\Delta R$ between the leading and subleading leptons & $x$ \\
$\Delta \eta_{\ell\ell}$ & Difference in pseudorapidity between the leading and subleading leptons & $\sqrt{x}$ \\
$m_{\ell\ell}$ & Invariant mass of the two leptons & $\log_{10}(x)$ \\
$\pt^{\ell\ell}$ & $\pt$ of the dilepton system & $\sqrt{x}$ \\
$\mt$ & Transverse mass of the dilepton system and $\MET$~\cite{STDM-2018-32} & $\sqrt{x}$ \\
$\mt^{o}$ & Projected transverse mass $\sqrt{(\pt^{\ell_1}+\pt^{\ell_2}+\MET)^2-(\vec{\pt}^{\ell_1}+\vec{\pt}^{\ell_2}+\vec{E}^\mathrm{miss}_{\mathrm{T}})^2}$~\cite{Barr:2011xt} & $\sqrt{x}$ \\
$\Delta R_{jj}$ & Distance $\Delta R$ between the leading and subleading jets & $x$ \\
$\Delta y_{jj}$ & Rapidiy difference between the leading and subleading jets & $x$ \\
$m_{jj}$ & Invariant mass of the two jets & $\log_{10}(x)$ \\
$\Delta \phi_{jj}$ & Difference in azimuthal angle between the leading and subleading jets & $x$ \\
$(\pt^{\ell_1} \cdot \pt^{\ell_2}) / (\pt^{j_1} \cdot \pt^{j_2})$ & $\pt$ ratio of the leptons and jets~\cite{Doroba:2012pd} & $\log_{10}(x + 0.02)$ \\
$\mathrm{min}(\Delta R_{\ell,j})$ & Minimal distance $\Delta R$ between the leptons and jets & $x$ \\
\hline
\end{tabular}}
\label{table:MVA_training_variables}
\end{table}

\begin{table}[t!bp]
\centering
\caption{Post-fit expected signal and background yields and observed data events in the SR. The yields are shown for the fit for the \sswwlx measurement in regions of the inclusive DNN score. The total uncertainties in the predicted yields are shown. The ``Conversions" category combines $V\gamma$ and charge misidentifcation background processes. The ``Other prompt'' category combines $ZZ$, $VVV$, $t\bar{t}V$, and $tZq$ background processes. The polarization vectors are defined in the $\ssww$ center-of-mass reference frame. \label{tab:yields}}
\begin{tabular}{c|c|c|c}
\hline
\hline
Process & \makecell{Region 0 to 0.3} & \makecell{Region 0.3 to 0.7} & \makecell{Region 0.7 to 1}  \\
\hline
$\sswwll$  &   \makecell{1.8 $\,\pm\,0.7$}  & \makecell{5.6 $\,\pm\,1.9$}  & \makecell{8.3 $\,\pm\,2.8$}  \\
$\sswwlt$  &  \makecell{6.0 $\,\pm\,2.5$}   &  \makecell{17.8 $\,\pm\,6.2$}  & \makecell{25.7 $\,\pm\,8.5$}  \\
$\sswwtt$  &  \makecell{18.4 $\,\pm\,4.4$}   &  \makecell{64.0 $\,\pm\,10.0$}  & \makecell{111.9 $\,\pm\,14.7$}  \\
QCD $\sswwjj$   &  \makecell{14.4 $\,\pm\,4.1$}  & \makecell{12.5 $\,\pm\,3.5$}   & \makecell{2.6 $\,\pm\,0.8$} \\
Int $\sswwjj$   & \makecell{2.3 $\,\pm\,0.1$}  &  \makecell{6.0 $\,\pm\,0.2$}  & \makecell{4.8 $\,\pm\,0.1$}  \\
QCD $\wzjj$  &  \makecell{33.3 $\,\pm\,2.6$}   & \makecell{20.3 $\,\pm\,1.6$} & \makecell{4.5 $\,\pm\,0.4$}  \\
EW $\wzjj$  & \makecell{3.3 $\,\pm\,0.1$}   & \makecell{6.2 $\,\pm\,0.2$} & \makecell{5.5 $\,\pm\,0.2$}  \\
Non-prompt    & \makecell{31.2 $\,\pm\,4.3$}      &   \makecell{20.2 $\,\pm\,2.9$}  & \makecell{10.4 $\,\pm\,3.1$}  \\
Conversions      & \makecell{14.6 $\,\pm\,3.7$} &            \makecell{6.2 $\,\pm\,1.8$}  & \makecell{1.6 $\,\pm\,0.5$}  \\
Other prompt    &     \makecell{3.7 $\,\pm\,0.7$} &      \makecell{2.6 $\,\pm\,0.6$}  & \makecell{0.8 $\,\pm\,0.2$}  \\
\hline
Total SM  & \makecell{129 $\,\pm\, 7$} & \makecell{161 $\,\pm\,10$} & \makecell{176 $\,\pm\,13$}  \\
\hline
Data   & \makecell{142} & \makecell{158} & \makecell{175} \\
\hline
\end{tabular}
\end{table}

\begin{table}[t!bp]
\caption{The fractional uncertainty of different components on the $\sigma \mathcal{B} (\sswwlx$) measurement. The contribution of a systematic uncertainty (uncertainty group) to the total uncertainty is evaluated by fixing the respective NP (NPs) to its (their) best-fit value(s), redoing the fit, and subtracting the uncertainties in the cross-section in quadrature. The procedure is implemented incrementally such that the sum in quadrature of the grouped systematic and statistical uncertainties corresponds to the total cross-section uncertainty by construction. Lepton calibration uncertainties encompass the effects of the calibration of lepton energy or momentum scale and resolution, as well as the lepton trigger, reconstruction, identification, and isolation efficiencies.  The ``Background modeling statistical'' category is related to the effect of a finite number of data events used for data-driven background estimates and of MC events. The polarization vectors are defined in the $\ssww$ center-of-mass reference frame. }
\begin{center}
\begin{tabular}{p{2ex}lc}
\toprule
\multicolumn{2}{l}{Source of uncertainty} & $\Delta \sigma / \sigma$ [\%] \\
\midrule
\multicolumn{2}{l}{Experimental} & \\
\midrule
& Lepton calibration                      & ${\hphantom{0}0.2}$ \\
& Jet energy and $E_{\mathrm{T}}^{\mathrm{miss}}$ scale and resolution & ${\hphantom{0}3.9}$ \\
& Pileup modeling                                                 & ${\hphantom{0}1.2}$ \\
& Background, charge misidentification and photon conversion        & ${\hphantom{0}0.5}$ \\
& Background, non-prompt                              & ${\hphantom{0}4.2}$ \\
& Background modeling statistical                                  & ${\hphantom{0}7.2}$ \\
& Luminosity                                                       & ${\hphantom{0}1.1}$ \\
\midrule
\multicolumn{2}{l}{Modeling} & \\
\midrule
& $W^{\pm}W^{\pm}jj$ EW + QCD uncertainties                                 & ${\hphantom{0}4.7}$ \\
& Background, WZ scale, PDFs \& $\alpha_s$                         & ${\hphantom{0}0.4}$ \\
& Background, WZ reweighting                                       & ${\hphantom{0}0.3}$ \\
& Small background normalizations                                  & ${\hphantom{0}1.1}$ \\
& Normalization factors                                            & ${\hphantom{0}2.5}$ \\
\midrule
\multicolumn{2}{l}{Experimental and modeling}                                & ${\hphantom{0}11.1}$ \\
\multicolumn{2}{l}{Data statistical}                               & ${\hphantom{0}32.2}$ \\
\midrule
\multicolumn{2}{l}{Total}                                          & ${\hphantom{0}34.1}$ \\
\bottomrule
\end{tabular}
\end{center}
\label{tab:unc}
\end{table}

\begin{table}[t!bp]
\centering
\caption{Measured and predicted fiducial cross sections for the $\sswwll$ and the $\sswwtx$, and the $\sswwlx$ and the $\sswwtt$ processes. The polarization vectors are defined in the initial-state parton--parton reference frame. The total measurement uncertainty is shown along with the statistical, modeling systematic and experimental systematic uncertainties. The modeling uncertainties include the theory uncertainties affecting the signal and background processes as well as the effect of a finite number of MC events and of data events used for data-driven background estimates. The predicted cross sections from \SHERPA[3] include the NLO EW corrections and the corrections to take into account the off-shell production and missing contributions. The predicted cross section uncertainties include the PDF, $\alpha_S$, and scale variations.}
\small
\begin{tabular}{cccc}
\toprule
Description & Predicted $\sigma \mathcal{B}$ (fb) & Measured $\sigma \mathcal{B}$ (fb) & Uncertainty breakdown (fb) \\
\midrule
\sswwll & \ensuremath{0.19 \pm 0.05} & \ensuremath{0.16 \pm 0.22 \text{\xspace(tot.)}} & \ensuremath{ \pm 0.21 \text{\xspace(stat.)} \pm 0.02 \text{\xspace(mod. syst.)} \pm 0.06 \text{\xspace(exp. syst.)}} \\
\sswwtx & \ensuremath{2.67 \pm 0.66}  & \ensuremath{3.40 \pm 0.35 \text{\xspace(tot.)}} & \ensuremath{ \pm 0.31 \text{\xspace(stat.)} \pm 0.08 \text{\xspace(mod. syst.)} \pm 0.16 \text{\xspace(exp. syst.)}} \\
\hline
\sswwlx & \ensuremath{1.24 \pm 0.31} & \ensuremath{0.84 \pm 0.37 \text{\xspace(tot.)}} & \ensuremath{ \pm 0.35 \text{\xspace(stat.)} \pm 0.05 \text{\xspace(mod. syst.)} \pm 0.11 \text{\xspace(exp. syst.)}} \\
\sswwtt & \ensuremath{1.62 \pm 0.39} & \ensuremath{2.46 \pm 0.37 \text{\xspace(tot.)}} & \ensuremath{ \pm 0.34 \text{\xspace(stat.)} \pm 0.06 \text{\xspace(mod. syst.)} \pm 0.14 \text{\xspace(exp. syst.)}} \\

\bottomrule
\end{tabular}
\label{tab:fiducialpp}
\vspace{10pt}
\end{table}

\clearpage

\printbibliography
\clearpage
 
\begin{flushleft}
\hypersetup{urlcolor=black}
{\Large The ATLAS Collaboration}

\bigskip

\AtlasOrcid[0000-0002-6665-4934]{G.~Aad}$^\textrm{\scriptsize 104}$,
\AtlasOrcid[0000-0001-7616-1554]{E.~Aakvaag}$^\textrm{\scriptsize 17}$,
\AtlasOrcid[0000-0002-5888-2734]{B.~Abbott}$^\textrm{\scriptsize 123}$,
\AtlasOrcid[0000-0002-0287-5869]{S.~Abdelhameed}$^\textrm{\scriptsize 119a}$,
\AtlasOrcid[0000-0002-1002-1652]{K.~Abeling}$^\textrm{\scriptsize 55}$,
\AtlasOrcid[0000-0001-5763-2760]{N.J.~Abicht}$^\textrm{\scriptsize 49}$,
\AtlasOrcid[0000-0002-8496-9294]{S.H.~Abidi}$^\textrm{\scriptsize 30}$,
\AtlasOrcid[0009-0003-6578-220X]{M.~Aboelela}$^\textrm{\scriptsize 45}$,
\AtlasOrcid[0000-0002-9987-2292]{A.~Aboulhorma}$^\textrm{\scriptsize 36e}$,
\AtlasOrcid[0000-0001-5329-6640]{H.~Abramowicz}$^\textrm{\scriptsize 157}$
% ...
\bigskip
\\

% ...
$^{17}$Department for Physics and Technology, University of Bergen, Bergen; Norway.\\
% ...
$^{30}$Physics Department, Brookhaven National Laboratory, Upton NY; United States of America.\\
% ...
$^{36}$$^{(a)}$Facult\'e des Sciences Ain Chock, Universit\'e Hassan II de Casablanca;$^{(b)}$Facult\'{e} des Sciences, Universit\'{e} Ibn-Tofail, K\'{e}nitra;$^{(c)}$Facult\'e des Sciences Semlalia, Universit\'e Cadi Ayyad, LPHEA-Marrakech;$^{(d)}$LPMR, Facult\'e des Sciences, Universit\'e Mohamed Premier, Oujda;$^{(e)}$Facult\'e des sciences, Universit\'e Mohammed V, Rabat;$^{(f)}$Institute of Applied Physics, Mohammed VI Polytechnic University, Ben Guerir; Morocco.\\
% ...
$^{45}$Physics Department, Southern Methodist University, Dallas TX; United States of America.\\
% ...
$^{49}$Fakult\"{a}t Physik , Technische Universit{\"a}t Dortmund, Dortmund; Germany.\\
% ...
$^{55}$II. Physikalisches Institut, Georg-August-Universit\"{a}t G\"ottingen, G\"ottingen; Germany.\\
% ...
$^{104}$CPPM, Aix-Marseille Universit\'e, CNRS/IN2P3, Marseille; France.\\
% ...
$^{119}$$^{(a)}$New York University Abu Dhabi, Abu Dhabi;$^{(b)}$United Arab Emirates University, Al Ain; United Arab Emirates.\\
% ...
$^{123}$Homer L. Dodge Department of Physics and Astronomy, University of Oklahoma, Norman OK; United States of America.\\
% ...
$^{157}$Raymond and Beverly Sackler School of Physics and Astronomy, Tel Aviv University, Tel Aviv; Israel.\\
% ...
$^{a}$ Also at Affiliated with an institute formerly covered by a cooperation agreement with CERN.\\
$^{b}$ Also at An-Najah National University, Nablus; Palestine.\\
$^{c}$ Also at Borough of Manhattan Community College, City University of New York, New York NY; United States of America.\\
$^{d}$ Also at Center for Interdisciplinary Research and Innovation (CIRI-AUTH), Thessaloniki; Greece.\\
$^{e}$ Also at Centre of Physics of the Universities of Minho and Porto (CF-UM-UP); Portugal.\\
$^{f}$ Also at CERN, Geneva; Switzerland.\\
% ...

\end{flushleft}


%

%
%

%

%
%
%
%

\end{document}